\documentclass[]{article}

\usepackage{a4wide}
\usepackage{epsf}
\usepackage{graphicx}
\usepackage[usenames]{color}
\usepackage[latin1]{inputenc}
\usepackage{hyperref}
\usepackage{float}

\usepackage{authblk}

\newcommand {\rad} {{\tt Radiance} }

\usepackage{listings,color}
\definecolor{verbgray}{gray}{0.9}

\hypersetup{pdfauthor={Peter Apian-Bennewitz},%
            pdftitle={Review of simulating four classes of window materials for daylighting with non-standard BSDF using simulation program Radiance},
            pdfsubject={Radiance materials modelling},%
            pdfkeywords={daylight simulation; window materials; BSDF; light scattering; light redirection; Radiance}
            pdfproducer={LaTeX},%
            pdfcreator={pdfLaTeX}
}

\usepackage{titlesec}
\titleformat{\section} {\normalfont\large\bfseries}{\thesection}{1em}{}
\titleformat{\subsection} {\normalfont\bfseries}{}{0em}{}
\titleformat{\subsubsection} {\normalfont\normalsize}{}{0em}{}

\definecolor{mygray}{rgb}{0.9,0.9,0.9}

\makeindex

\begin{document}

\setlength{\voffset}{-1.0in}
\setlength{\topmargin}{1.0cm}
\setlength{\headheight}{2.0ex}
\setlength{\headsep}{1.0cm}
\setlength{\footskip}{1.0cm}
\setlength{\textheight}{24.0cm}

\setlength{\parindent}{0cm}

\title{Review of simulating four classes of window materials for daylighting with non-standard BSDF using the simulation program \rad}

\author{Peter Apian-Bennewitz
\footnote{contact email: papers@pab.eu}}
\affil{pab advanced technologies Ltd}

\date{dated: \today}


\maketitle

\section{introduction}

From a practical point of view, daylighting in architecture uses the visible spectrum of solar energy directly and efficiently.
It offers potential to decrease the energy consumption of a building and enhance the aesthetics of indoor spaces.
Optimisation of daylighting can be roughly divided into two approaches, or a combination of the two in one project: First, the shape of the
building can be optimised for a specific location and climate, using traditional materials. This is the classical factor of climate and
local conditions on architectural shapes \cite{lechner:91}.
Secondly, the use of new window materials with complex light-scattering and -redirection offers the potential to enhance daylighting with fewer
requirements on the building shape. This has been the subject of research over a number of years \cite{wittwer:90}, \cite{compagnon:93},
\cite{joule2-rep}, \cite{jelle:2011}.

From a physics point of view, the core concept of the physics and numerics of light-scattering and -redirection is the
{\em bidirectional scatter distribution function} (BSDF) \cite{nicodemus-1977-geometric}, \cite{stover:90}. Lighting properties of window
materials, whether structured on a macroscopic scale (embedded mini-louvers between panes) or microscopic scale (e.g. holografic elements),
can be described with the BSDF concept fittingly. However, a detailed description of the relations between the physics of the BSDF, the possible
advantages of a material for daylighting and the numerical details of algorithms to simulate it would exceed this text and is left to a
specific further paper. 

Tools for simulation of daylighting have to predict accurate lighting levels in real-world applications, including materials with complex
scattering and redirecting properties. This sets them apart from general rendering in computer graphics, where materials predominantly have
simpler optical properties and ''just have to {\em look} right''.


Daylight simulations are frequently using the software \rad, \cite{galasiu:2008} found a market share greater 50\% market in their 2006 survey.
\rad has been developed since 1990 by Greg Ward, then at Lawrence Berkeley Lab, CA, USA \cite{ward-1992-irradiance}.
A number of other raytracing programs are established in the optics industry (e.g. for lens design) and other programs, some based on
less general Radiosity algorithms, have been commercially available for daylighting. \rad is still in active use, either a stand-alone tool, in
open-source front-ends \cite{openstudio:2013} or in commercial programs \cite{relux:2010}, \cite{rayfront:2008}.
The core calculation engine of \rad, having been designed for day- and artificial lighting from the start, offers a number of advantages,
for example: support for non-diffuse surfaces, efficient algorithms with comparatively low memory footprint and source code availability.
%
Historically, not all ''legacy'' window materials in \rad are supported in all lighting situations, due to limits of the built-in algorithms.
Even recent papers are not specifically detailed about this \cite{reinhart-trans:2006}.

Some older advanced material models of \rad have been around for a long time, without much documentation, but can still be a well adapted choice 
to model advanced materials today. Furthermore, new features have been introduced in the last 2 years (BSDF material, genBSDF \cite{ward:2011}).
Correlations between newly added materials and older special types do not seem to have been documented.

This review tries to give a practical guide to all currently implemented window material models in \rad , while only briefly mentioning the
details of the Physics of BSDFs or details of numerical algorithms.
Not included are comparisons of BSDF models and measured data for window materials \cite{bme:2011}, since it would fill about the same
number of pages. The same applies to further processing steps like glare analysis and annual calculations. 

Four classes of models for window materials are presented, based on light {\em redirection} and {\em scattering}:\\
Light {\em redirection}, as defined here, is an ideal redirection of incident light into one or more discrete directions. Their
bidirectional-scatter-distribution-function (BSDF, see \ref{BSDFdef}, consists of a sum of delta-peaks \cite{diracdelta}, each corresponding
to a redirected direction. The BSDF is zero in between, since there is no further scattering into other directions.\\
Light {\em scattering} refers, in this context, to light scattered around a pronounced direction, not necessarily limited to the forward direction.

\begin {figure} [hbt]
\begin {center}
\includegraphics[width=10cm]{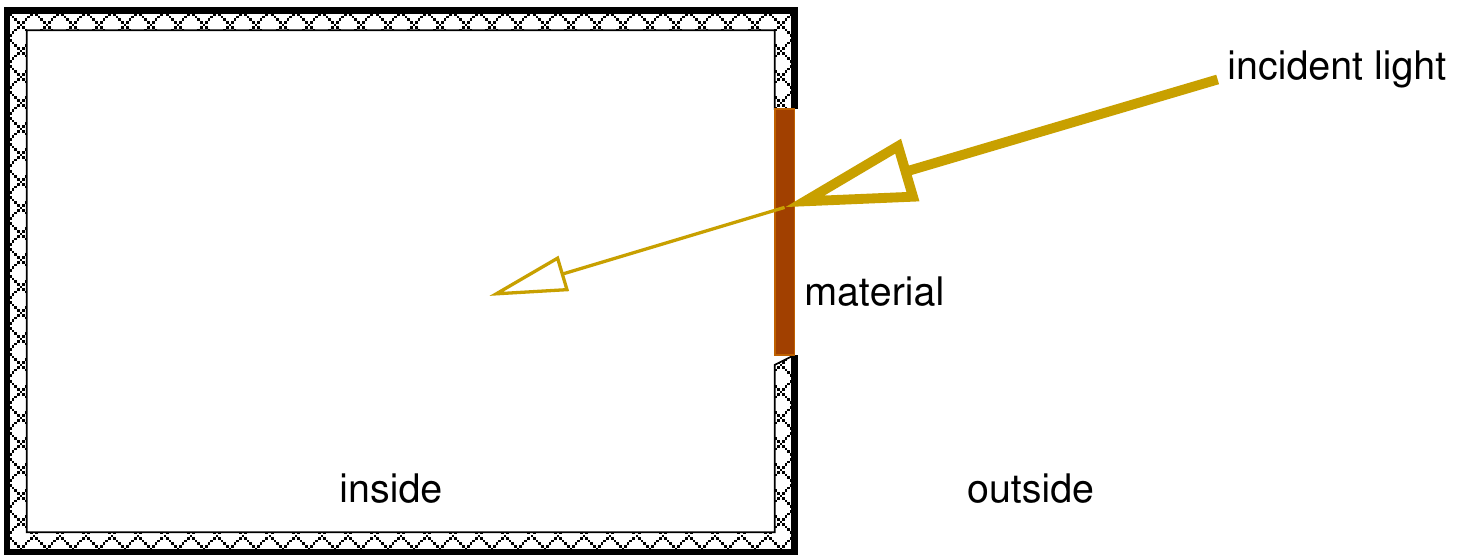}
\caption {\label{non-redir-non-scat}Light paths for clear, non-redirecting, non-scattering window elements}
\end {center}
\end {figure}

\begin {figure} [hbt]
\begin {center}
\includegraphics[width=10cm]{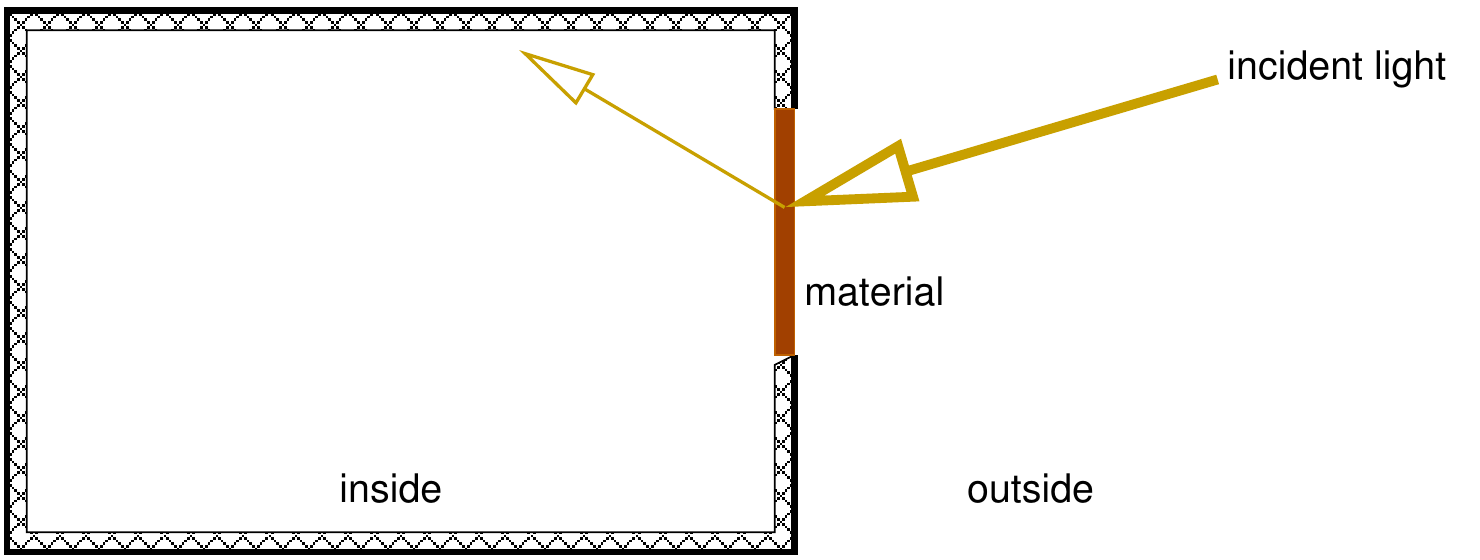}
\caption {\label{redir-non-scat}Light paths for redirecting, non-scattering window elements}
\end {center}
\end {figure}

\begin {figure} [hbt]
\begin {center}
\includegraphics[width=10cm]{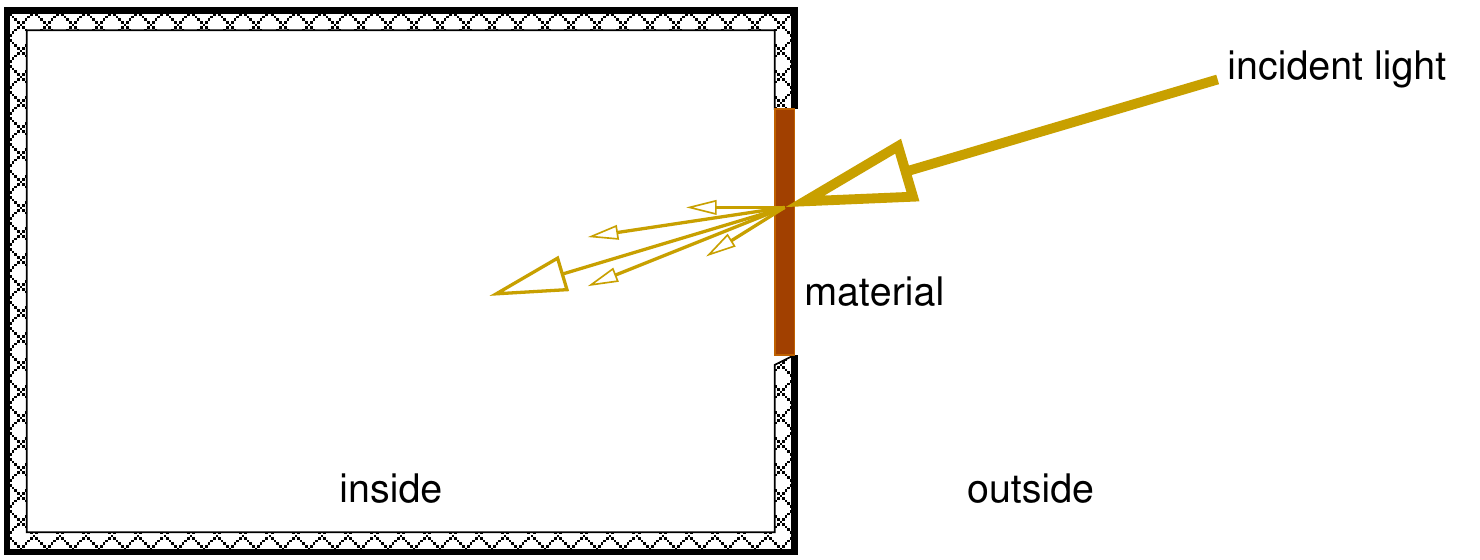}
\caption {\label{non-redir-scat}Light paths for non-redirecting, scattering window elements}
\end {center}
\end {figure}

\begin {figure} [H]
\begin {center}
\includegraphics[width=10cm]{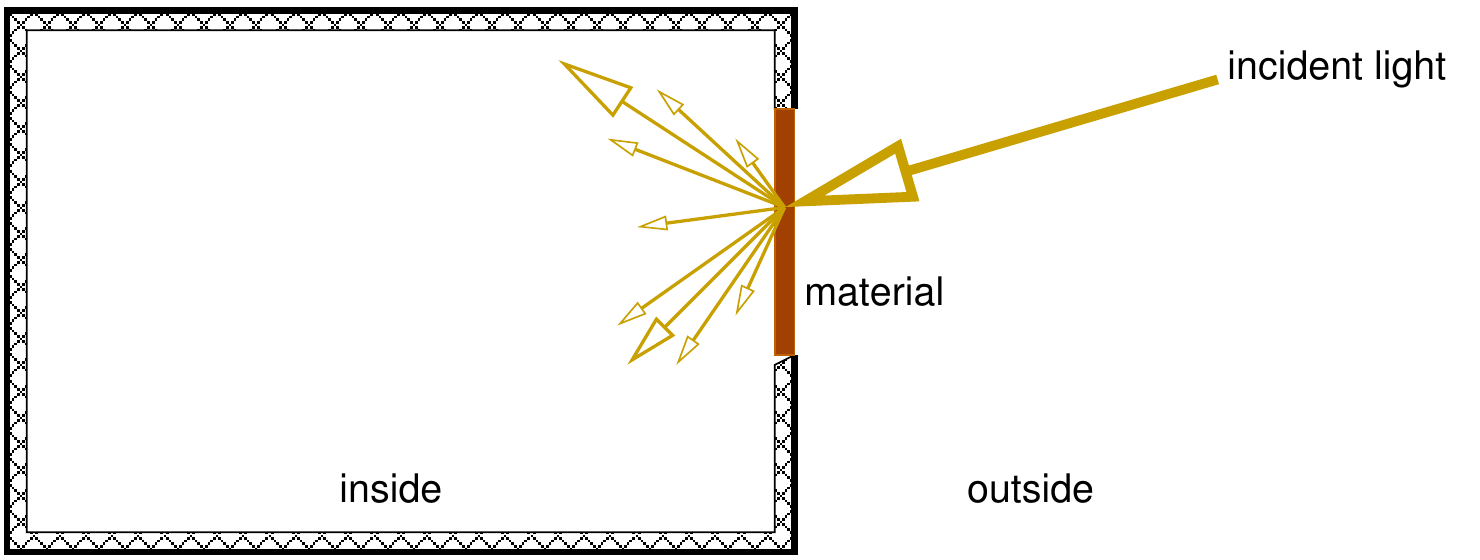}
\caption {\label{redir-scat}Light paths for redirecting, scattering window elements}
\end {center}
\end {figure}

This leads to four combinations:
Non-redirecting+non-scattering (Fig.~\ref{non-redir-non-scat}, e.g. clear glass),
redirecting\-non-scattering (Fig.~\ref{redir-non-scat}, e.g. ideal prismatic material) and
Non-redirecting+scattering (Fig.~\ref{non-redir-scat}, e.g. frosted glass),
redirecting+scattering (Fig.~\ref{redir-scat}, e.g. multi-walled extrusion materials).
An interesting fact is that these groups correlate with the numerical algorithms and are helpful to show their limits.

The discussion is of specific relevance to users of the simulation program \rad. Users of other programs are invited to compare results of
their programs to \rad, detailed feedback is welcome.
Knowledge-wise this text starts somewhat in the middle of a typical learning curve with \rad: It assumes some familiarity with
its overall concept, sky models, rendering parameters, plus some basic idea about light scattering at materials.


\subsection{Test scene}

\begin {figure} [hbt]
\begin {center}
\includegraphics[width=10cm]{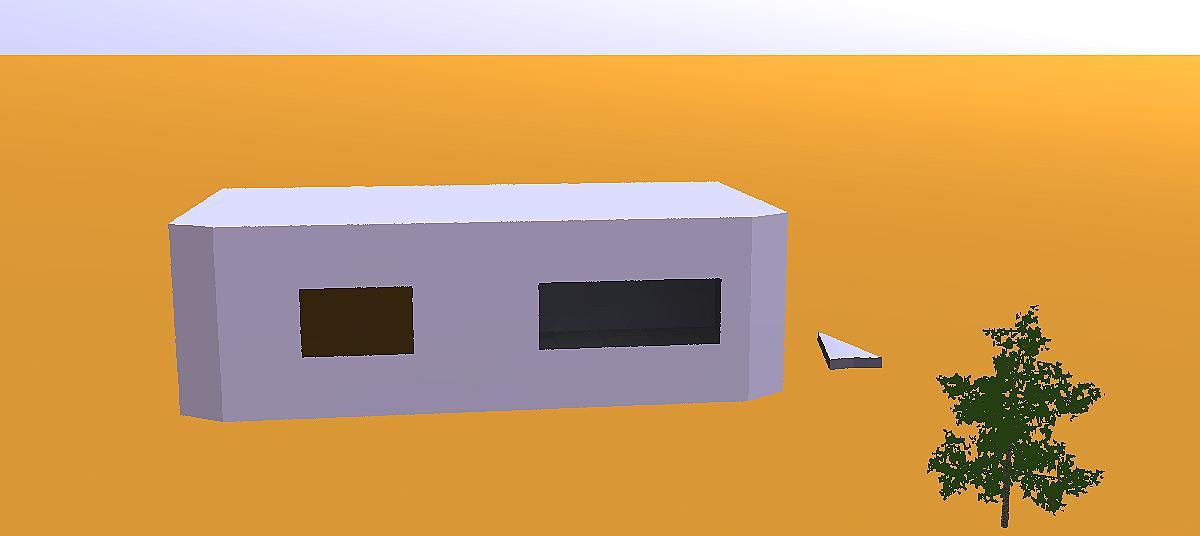}
\caption {\label{outside}Outside view of test scene, reference window on left, test-window on right }
\end {center}
\end {figure}

The test-room (Fig.~\ref{outside}) has a reference window, test-window (Fig.~\ref{outside}) and a tree for visual reference.
Location is 48.0N (Freiburg), April 12th, 10:30am,  the outside marker points North.

This model is kept simple and does not include a geometrical plane around the building, since it would make understanding the light
transport at the window unnecessarily complex. It uses a solid angle for the light coming from below the horizon and a solid angle for the
light coming from the non-sun part of the sky. Consequently, there are no shadows of the building on the surrounding ground.  The sky solid
angle is purposely exaggerated bluish and the ground solid angle is exaggerated brownish to mark the source of the
\index{horizon}
sky light.
Needless to say, today's simulation programs, including \rad{}, can handle more geometrically complex scenes easily.

\pagebreak[6]
\subsection{Terminology of \rad and algorithms for light simulation}

\begin{description}

\item[eye rays]
\label{eyerays}
Calculation of light transported from a point in the scene towards to view-point or eye. These rays render the pixels of an image.
Technically a ray is traced from eye towards the scene, new rays may be spawned at the first intersection with the scene geometry.

\item[direct rays, direct calculation]
\label{directrays}
Calculation of energy transported in a straight line from a light source to the receiving point.
In day-lighting, this refers to light travelling from the sun directly to an indoor point.
Technically a ray is traced from the receiving point towards each light source and checked for intermediate surfaces.\\
These calculations do not find alternative paths between source and receiver, if that paths is not a straight line. Examples
include: reflectors, lenses, light-pipes, prismatic glazings etc, see \ref{1-4-dielectric2-texture}.

\item[ambient rays]
\label{ambientrays}
Recursive calculation of light transported via other surfaces in the scene to the receiving point.
Technically a bunch of rays are traced from the receiving point towards random directions in the hemisphere.
Since ambient calculation is stochastic, results include random noise.
See \ref{0-2-glass2} for an example where the solid angle of the transmissive part of a window from an indoor point is small, requiring a high
number of ambient rays to sample it.

\item[photon map]
\label{pmap}
Forward raytracing of light from sources towards surfaces in the scene, including reflection and refraction. Works best for tracing direct
sunlight through complex fenestration. Available for \rad as an extra package \cite{schregle:04}, which is used as a preprocessor.  Since it
uses its own set of features and parameters, the examples do not include it here for reasons of overall volume of this text.
See also the more general {\em Metropolis Algorithm} in \cite{veach:97}.

\end{description}

\subsection{Terminology of light scattering}

\begin{description}

\item[BSDF]
\label{BSDFdef}
\index{ BSDF ! definition }
The Bidirectional-Scatter-Distribution-Function describes the interaction between light and material
within the {\em radiometric} framework of Physics \cite[part 7 ''Radiometry and Photometry'']{handbook_vol2}, \cite{stover:90}.
A short summary of BSDF properties: It depends on two directions: The incident direction $\vec{x}_{in}$ and the outgoing
direction $\vec{x}_{out}$, plus additional parameters, such as the position on sample (see example in \ref{0-2-glass2}) or wavelength.\\
Values of the BSDF are always positive, but can be larger than 1 . The integral over the outgoing hemisphere is maximal 1 for any incident
direction, due to energy conservation:\\ $\int^{hemisphere} f(\vec{x})\, \cos\theta_{o}\, d\vec{x} \le 1$, where $\theta_{o}$ is the angle between
surface normal and $\vec{x}$.\\
The BSDF is symmetric under exchange of $\vec{x}_{in}, \vec{x}_{out}$, due to the reversibility of light paths in Radiometry.
A constant BSDF describes a perfect diffusing sample. A BSDF with only a single Dirac-delta peak \cite{diracdelta} represents a
non-scattering sample \cite{stover:90}.
\index{Lambertian scattering} \index{scattering ! Lambertian} \index{ BSDF ! delta peak }

\item[Fresnel formula]
\label{fresnel}
The Fresnel equations describe the transmission and reflection of electromagnetic waves (including light) at a planar boundary between two
dielectric materials (e.g. at the surface between glass and air) \cite{jackson:75}.

\end{description}

\subsection{Terminology specific to this article}

\begin{description}

\item[non-sun sky]
A model of the radiance distribution of the sky, excluding the solid angle of the geometric sun shape, including the circum-solar radiation.
The exact distribution depends on the sky model (CIE, Perez, etc) and outside shading, e.g. by overhangs or neighbour buildings. However, the
numerical interactions between material model and sky model, described here, are valid for any sky model.
Technically, \rad samples the ''non-sun sky'' using ambient rays, whereas the sun itself is taken into account with direct rays.

\item[sunny sky]
A model of the sky distribution including the solid angle and radiance of the sun.

\end{description}


%
%
\newpage
\section{non-redirecting, non-scattering materials}
Incident light is neither scattered nor redirected at the window material. The intensity (more precisely the radiance) is scaled,
by implicit laws of Physics or user-supplied functions when a ray crosses through the window material.

This includes, for example:
\begin{itemize}
\item	Ideal glass pane with reflection and transmission by Fresnel's formula 
\item	sputtered layers and coatings, as long as scattering is not taken into account
\item	angular selective materials: micro-structured glass panes that show special angular dependency of reflection or transmission with incident angle,
	as long as they are modelled as reflective or absorptive, but not scattering
\item	Glass panes with adhesive foils as sun-shades, modelled without scattering
\end{itemize}
The BSDF of such materials consists of one ideal delta-peak at the forward direction of the unscattered beam. Any angular dependency, either
implicitly modelled by Fresnel formulas or by a user supplied function, scales the height of the BSDF peak, but not its position or shape.

Since the BSDF is symmetric in the incident and outgoing direction, a material that transmits the incoming visible spectrum unscattered will also
feature an unscattered view from the inside to the outside along the same direction. For this class of materials the view to the outside is clear,
but may be dimmed or tinted.

\vspace{1.0cm}

\fbox{\parbox[t][15em][c]{\textwidth}{
Note on the format of the following pages: For ease of reading, reference examples are compiled to summarise one material and sky condition
on one page, grouped by the four material classes.
\\
Each class is preceded by a description of the scattering characteristics for each group (like this page), listing a selection of materials
that are covered by the class.\\
Each page consists of: A description of the window type, the simulation result including false-colour contour lines of
indoor illuminance, the material description of the test window, the sky model (generated by \rad program {\tt gensky}) and the parameters used for
the simulation program {\tt rpict}.\\
Details are specified for this material, advantages (labelled {\bf Pro:}) and drawbacks ({\bf Con.:}) are given.\\
Note that the second window on the right side in each image is a 5\% transmitting glass, used as reference. Due to the low
transmission, the outside view is dimmed and the influence on inside illumination is negligible.
}}

\newpage

\setlength{\parskip}{0.8ex}
\subsection{The simplest case: Open window in the wall, without sun
}

\begin{tabbing}
Objective:	\={\bf Model room with a hole in the wall, non-sun sky
}\\
\>\\
Scene:		\>{\bf Walls modelled as opaque polygons, window area is left open
}\\
\end{tabbing}

\vspace*{-2ex}
\includegraphics[width=15cm]{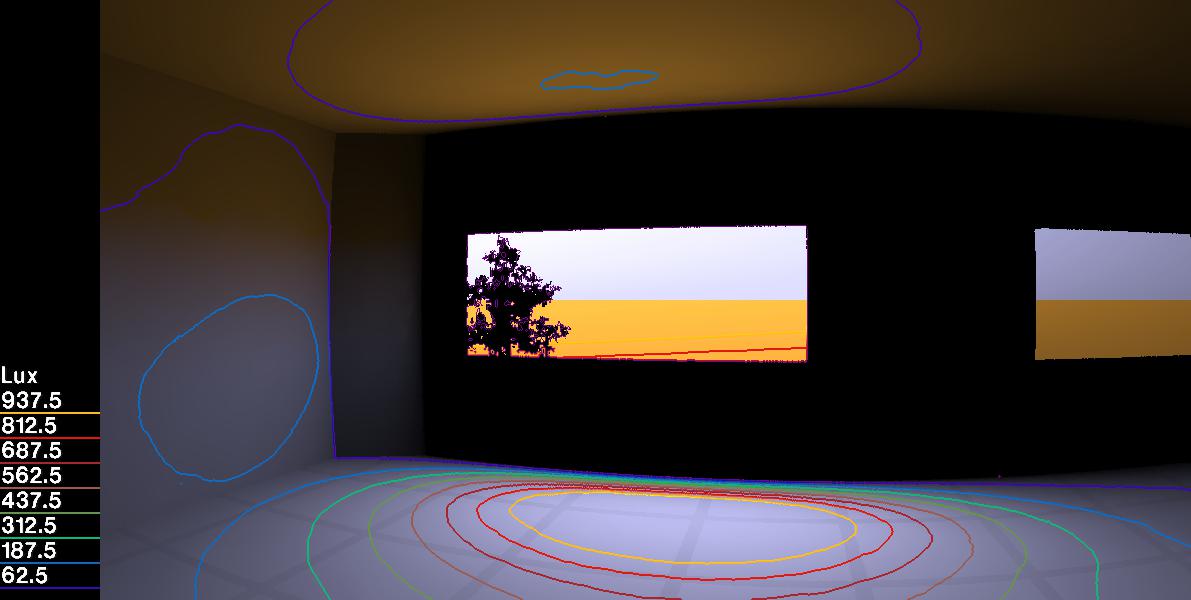}
\label{0-0-void1}

\lstset{ 
	backgroundcolor=\color{mygray},
	basicstyle=\ttfamily\footnotesize,
	breaklines=true,
	prebreak=\textbackslash,
	breakatwhitespace=true,
	showspaces=false,
	frame=lines,
}
\vspace*{-2ex}
\begin{minipage}[t]{8cm}		
\begin{minipage}[t]{7.4cm}		
\begin{lstlisting}
gensky 04 12 10.5CEST -a 48 -o -7.5 -i 
\end{lstlisting}
\end{minipage}\\
\begin{minipage}[t]{7.4cm}		
\begin{lstlisting}
rpict -ab 1 -ad 20000 -ar 512 
\end{lstlisting}
\end{minipage}
\end{minipage}
\begin{minipage}[t]{7cm}		
\begin{lstlisting}
void
\end{lstlisting}
\end{minipage}

\index{ glass pane ! none}
The contribution of the sky at an interior point is calculated by ambient rays (see \ref{ambientrays}), which are sent from an
interior point out in all directions. Some leave through the window and hit the sky or ground.\\
Eye rays (\ref{eyerays}) are sent from the eye position of an image towards the scene, forming the image of the window itself
in the rendering. At the window, they are handled similar to ambient rays, pass through the window and hit the outside \emph{sky glow} and
\emph{ground glow} domes.

Notes on model: In these example renderings, all wall, floor and ceiling materials are modelled as diffuse, grey surfaces.
The floor has a grid pattern for geometric reference.
The sky dome is purposely exaggerated bluish and the ground dome is purposely exaggerate brownish to mark the source of the ambient light.

Note the bluish tint of  light from the sky on the floor, and the reddish tint of light from the ground on the ceiling.
Also note the asymmetry of irradiance levels on the floor due to the asymmetric morning sky distribution.

Unless otherwise noted, all renderings use an \emph{ambient bounce} of 1 . Full simulations of interior light levels would use a higher
value, but a value of 1 shows the influence by the window and sky most clearly.

The window on the right edge of the image serves as reference in these images and has a glass of 5\% transmission.
The difference in colour in the visible image of the two windows is due to {\em tone mapping}, the process of converting radiance and
luminance values of a simulation (high-dynamic-range images, HDR) to low-dynamic-range images for publishing.

{\bf Pro:} This material is fully supported by all light calculations in \rad. It is included here for reference of illuminance levels compared to the other materials.

{\bf Con.:} 
A very simple example.

\vfill

\newpage

\setlength{\parskip}{0.8ex}
\subsection{Simple glass window, without sun
}

\begin{tabbing}
Objective:	\={\bf Model room with a single glass pane, non-sun sky
}\\
\>\\
Scene:		\>{\bf Window is material type {\tt glass}
}\\
\end{tabbing}

\vspace*{-2ex}
\includegraphics[width=15cm]{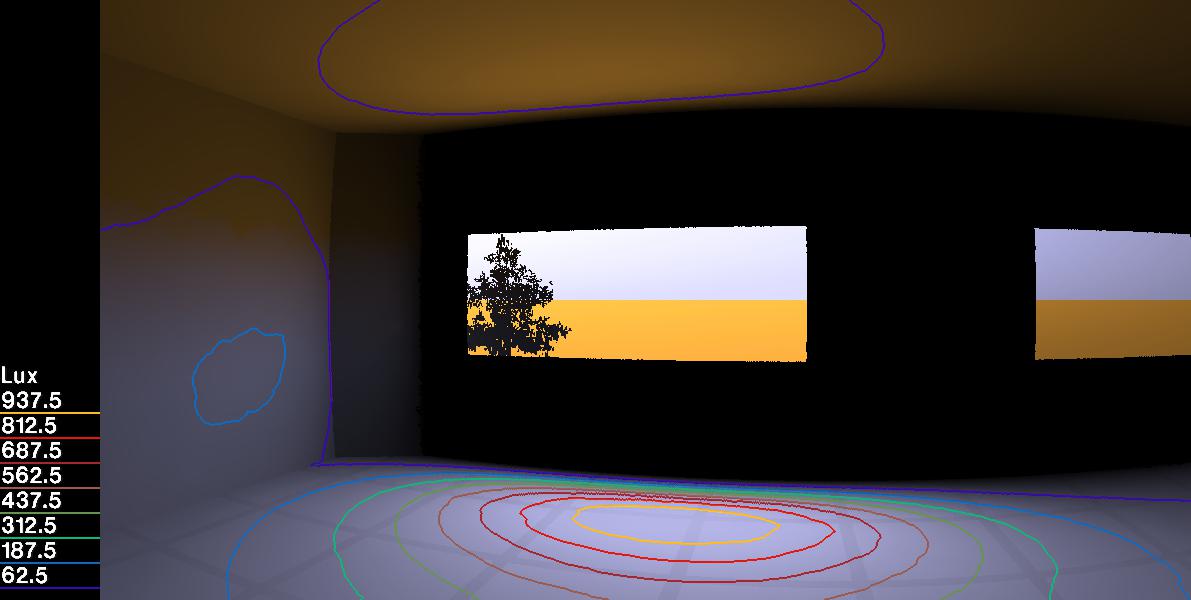}
\label{0-2-glass1}

\lstset{ 
	backgroundcolor=\color{mygray},
	basicstyle=\ttfamily\footnotesize,
	breaklines=true,
	prebreak=\textbackslash,
	breakatwhitespace=true,
	showspaces=false,
	frame=lines,
}
\vspace*{-2ex}
\begin{minipage}[t]{8cm}		
\begin{minipage}[t]{7.4cm}		
\begin{lstlisting}
gensky 04 12 10.5CEST -a 48 -o -7.5 -i 
\end{lstlisting}
\end{minipage}\\
\begin{minipage}[t]{7.4cm}		
\begin{lstlisting}
rpict -ab 1 -ad 20000 -ar 512 
\end{lstlisting}
\end{minipage}
\end{minipage}
\begin{minipage}[t]{7cm}		
\begin{lstlisting}
void glass testmat
0
0
3 0.96 0.96 0.96

\end{lstlisting}
\end{minipage}

\index{glass pane ! clear ! thin }
The contribution of the sky at an interior point is calculated by ambient rays, which hit the window, are scaled by the transmission formula and
continue towards the sky or ground. Eye rays are handled the same way as ambient rays.

Compared with \ref{0-0-void1} the inside illuminance levels are lower, due to the reflection of light at the window (\emph{Fresnel formula}.
An optional fourth parameter sets the index of refraction, the default is $n=1.52$.
Transmittance of light inside the window material is given by the first three parameters.

Compared to the {\tt dielectric} material in \ref{1-1-dielectric1}, this model of a glass pane is infinitely thin. It combines the Fresnel
coefficients of front- and back-side, plus the sum of multiple internal reflections between the front and back surface and absorption inside the
material into one formula. The advantage is a performance gain compared to tracing reflections inside the material explicitly
(see \ref{1-1-dielectric1}). It also helps geometric modelling, since the glass pane is modelled as a single surface.

The window on the right edge of the image serves as reference in these images and has a glass of 5\% transmittance.
For the difference in colour of the two windows see \ref{0-0-void1}.

{\bf Pro:} This material is fully supported by all light calculations in \rad

{\bf Con.:} 
A fairly simple example. Does not model thick glass volumes (see \ref{1-1-dielectric1}).

\vfill

\newpage

\setlength{\parskip}{0.8ex}
\subsection{Glass window, partially tinted, without sun 
}

\begin{tabbing}
Objective:	\={\bf Model room with a single glass pane of variable transmission, non-sun sky
}\\
\>\\
Scene:		\>{\bf Window is material type {\tt glass} combined with {\tt brightfunc}
}\\
\end{tabbing}

\vspace*{-2ex}
\includegraphics[width=15cm]{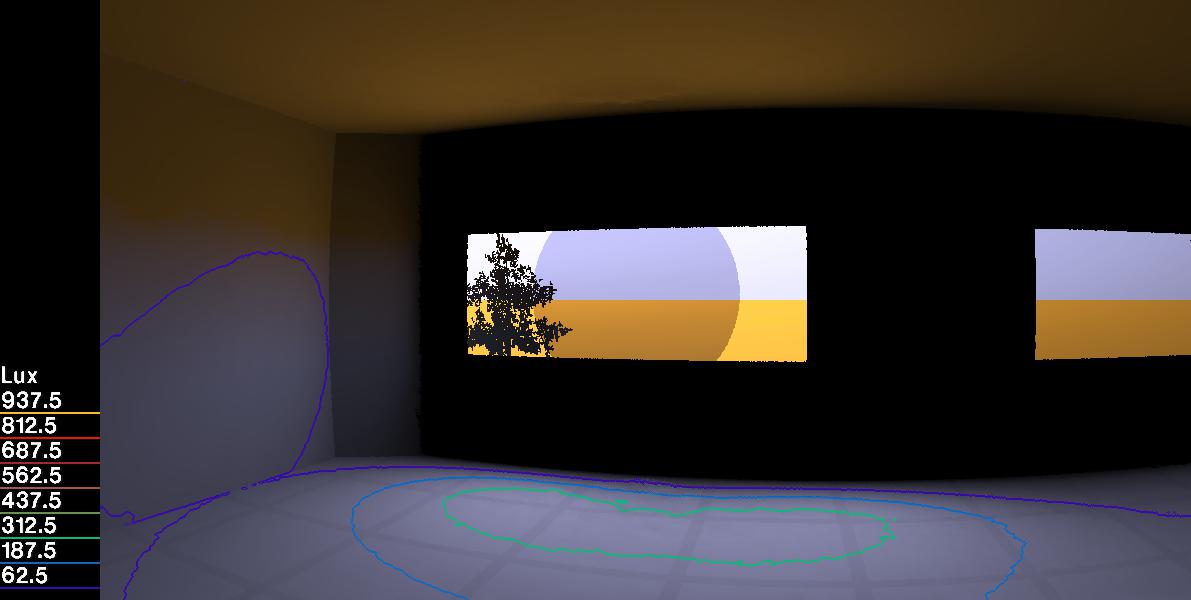}
\label{0-2-glass2}

\lstset{ 
	backgroundcolor=\color{mygray},
	basicstyle=\ttfamily\footnotesize,
	breaklines=true,
	prebreak=\textbackslash,
	breakatwhitespace=true,
	showspaces=false,
	frame=lines,
}
\vspace*{-2ex}
\begin{minipage}[t]{8cm}		
\begin{minipage}[t]{7.4cm}		
\begin{lstlisting}
gensky 04 12 10.5CEST -a 48 -o -7.5 -i 
\end{lstlisting}
\end{minipage}\\
\begin{minipage}[t]{7.4cm}		
\begin{lstlisting}
rpict -ab 1 -ad 20000 -ar 512 
\end{lstlisting}
\end{minipage}
\end{minipage}
\begin{minipage}[t]{7cm}		
\begin{lstlisting}
void brightfunc bf
2 val file1.cal
0
0
bf glass testmat
0
0
3 0.96 0.96 0.96

\end{lstlisting}
\end{minipage}

\index{ glass pane ! clear ! thin }
\index{ glass pane ! partially tinted }
One step more complex than \ref{0-2-glass1}: When passing through the window, ray values are multiplied by a user defined function.
In this case it is defined in a text file {\tt file1.cal}:
\begin{lstlisting}
SQR(x)=x*x;
r= sqrt( SQR(Px-7.2) + SQR(Pz-1.55));
val = if( r - 0.9, 1, 0.05 );
\end{lstlisting}
Transmission is scaled down by a factor of 0.05 in a circle with 0.9m radius around coordinates x=7.2 z=1.55 .
For variables inside function files, like the position vector ({\tt Px,Py,Pz}), see file {\tt rayinit.cal}, included in the \rad package.
An angular dependency can be modelled as well, by dependency on the D-vector ({\tt Dx,Dy,Dz}), which gives the ray direction
through the material.\footnote{See \ref{3-2-transfunc1-sun} for the more complex case of function files for scattering materials.}
Note the different indoor illuminance levels, compared to \ref{0-2-glass1}.

{\bf Pro:} This material is fully supported by all light calculations in \rad.
Function files allow complex patterns which are ideal for parametric studies and are more numerically efficient than geometric models for finely
structured window materials.

{\bf Con.:} 
The function scales transmission and reflection only, it does not model scattering. A selectively etched glass pane, or fabrics for sunshades
can {\em not} be modelled this way. See \ref{2-1-mixfunc1} in how to model selectively etched glass.
\index{ fabrics }\\
Since the resulting transmissive part of the window is smaller, a higher {\tt -ad} parameter may be required to reduce the slightly higher random noise
in these ambient calculations.

\vfill

%
%
\section{redirecting, non-scattering materials}
Incident light onto a window of this type is redirected without scattering. Additionally it may be scaled by absorption or reflection coefficients.

This includes for example:
\begin{itemize}
\item	ideal louvre systems with flat glass or metal mirrors
\item	ideal prismatic glazings, modelled without scattering and chromatic aberrations
\end{itemize}
The BSDF of such materials consists of ideal delta-peaks. For some materials, the position of a number of delta-peaks can be given by function
file ({\tt prism1},{\tt prism2}), with other materials the position is given by a geometrical volume of this material ({\tt dielectric}).\\

Since the BSDF is symmetric in the incident and outgoing direction, a non-scattering material redirects the incoming light unscattered, and
it also redirects the view of the outside, resulting in a clear but shifted or distorted view.

The assumption of an ideal redirection without scattering is more adequate for pre-tests or proof-of-concept studies, than for actual project
work. Numerically the algorithms are faster than those which include scattering.

Additionally, complex geometries can be pre-processed using the \rad tools {\tt mkillum} \cite{radiance-man} or the newer and more powerful
{\tt genBSDF} \cite{ward:2011}.

\newpage

\setlength{\parskip}{0.8ex}
\subsection{Solid glass window, sky without sun
}

\begin{tabbing}
Objective:	\={\bf Model room with a solid block of glass as window, non-sun sky
}\\
\>\\
Scene:		\>{\bf Window is material type {\tt dielectric}, non-sun sky
}\\
\end{tabbing}

\vspace*{-2ex}
\includegraphics[width=15cm]{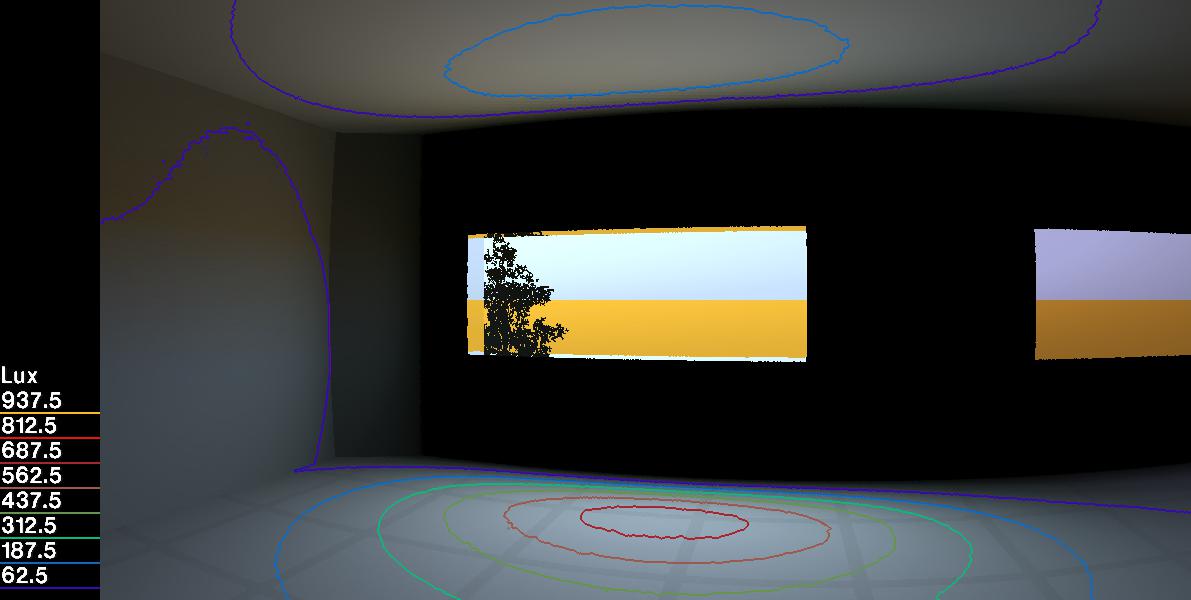}
\label{1-1-dielectric1}

\lstset{ 
	backgroundcolor=\color{mygray},
	basicstyle=\ttfamily\footnotesize,
	breaklines=true,
	prebreak=\textbackslash,
	breakatwhitespace=true,
	showspaces=false,
	frame=lines,
}
\vspace*{-2ex}
\begin{minipage}[t]{8cm}		
\begin{minipage}[t]{7.4cm}		
\begin{lstlisting}
gensky 04 12 10.5CEST -a 48 -o -7.5 -i 
\end{lstlisting}
\end{minipage}\\
\begin{minipage}[t]{7.4cm}		
\begin{lstlisting}
rpict -ab 1 -ad 20000 -as 5000 -ar 1024 -aa 0.05 
\end{lstlisting}
\end{minipage}
\end{minipage}
\begin{minipage}[t]{7cm}		
\begin{lstlisting}
void dielectric testmat
0
0
5 0.5 0.85 0.5 1.4 0

\end{lstlisting}
\end{minipage}

Models a window pane with finite thickness, whereas case \ref{0-2-glass1} models it infinitely thin. 
To demonstrate this difference, the dielectric window has a thickness of 0.5m, it is a solid block of glass.
The first three parameters (RGB) specify {\em internal transmittance} per unit length. In case of optical glass with an exemplary transmittance between 0.98 to
0.995 for a 25mm sample, the values range from 0.45 to 0.82 for a 1000mm sample. Since this model uses the length unit Meter, these
values are approximately correct, modulated by the actual glass type used. A green tint usually depends on the amount of iron in the material.\\
Additionally to the internal transmittance, the reflections at the front and rear surfaces depend on the index-of-refraction (fourth
parameter) relative to air (index 1) and are calculated according to Fresnels formula.\\
Note the total reflection at the edges of the glass block and the increased illuminance at the ceiling due to reflection of the non-sun
sky at the lower edge of the solid glass block.\\
The same features and limits apply to the more general material model {\tt interface} of two arbitrary dielectrics.
\index{ interface ! radiance material }
'

{\bf Pro:} Supported with eye and ambient rays, including multiple internal reflections.
For the contribution of the non-sun part of the sky, the standard backward raytracing has a higher computing efficiency than forward
raytracing (e.g. using the {\em photon map} algorithm).

{\bf Con.:} 
No support for refraction and reflection of direct sun (see \ref{1-2-dielectric-direct}).
No scattering is considered at front or back side or inside the dielectric. Wavelength dependent refraction is only marginally supported.\\
A geometry of {\tt dielectric} material must be a closed volume with correct surface orientations for the raytracing to work correctly.\\ 
To calculate the reflections of the non-sun sky at the glass edges in this example, a higher {\tt ad} parameter is needed to reduce noise.
Tracing multiple internal reflections takes longer to calculate than using the implicit formula for a thin layer of dielectric material in
{\tt glass} (\ref{0-2-glass1}).

\vfill

\newpage

\setlength{\parskip}{0.8ex}
\subsection{Solid glass window, sky with sun
}

\begin{tabbing}
Objective:	\={\bf Model room with a solid block of glass as window, sunny sky
}\\
\>\\
Scene:		\>{\bf Window is material type {\tt dielectric}, sky with sun
}\\
\end{tabbing}

\vspace*{-2ex}
\includegraphics[width=15cm]{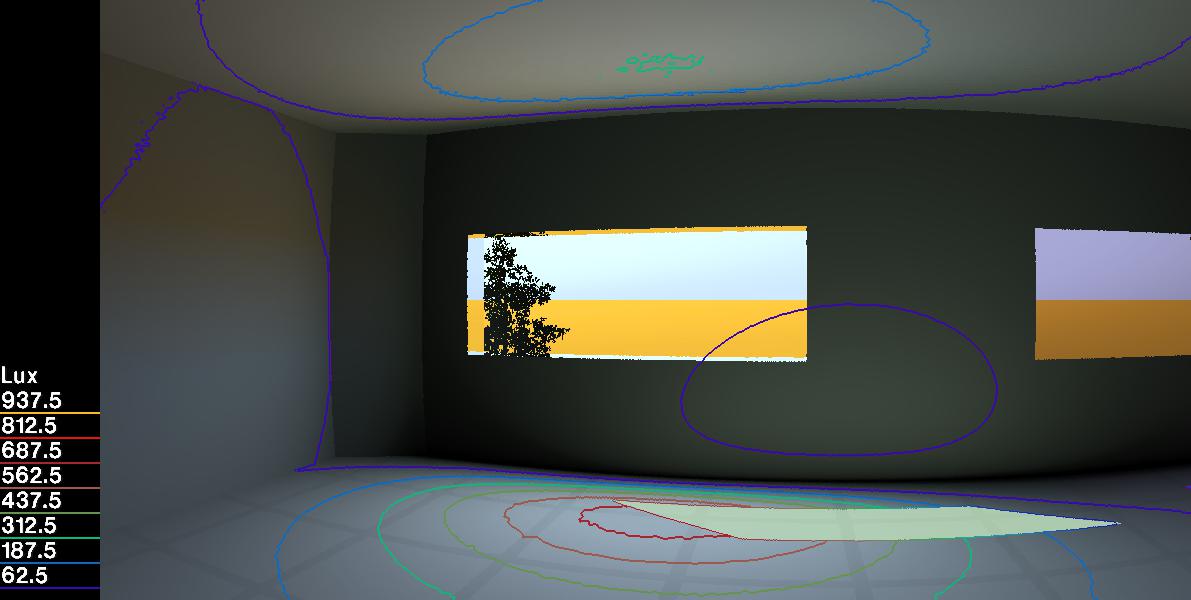}
\label{1-2-dielectric-direct}

\lstset{ 
	backgroundcolor=\color{mygray},
	basicstyle=\ttfamily\footnotesize,
	breaklines=true,
	prebreak=\textbackslash,
	breakatwhitespace=true,
	showspaces=false,
	frame=lines,
}
\vspace*{-2ex}
\begin{minipage}[t]{8cm}		
\begin{minipage}[t]{7.4cm}		
\begin{lstlisting}
gensky 04 12 10.5CEST -a 48 -o -7.5 +i 
\end{lstlisting}
\end{minipage}\\
\begin{minipage}[t]{7.4cm}		
\begin{lstlisting}
rpict -ab 1 -ad 20000 -as 5000 -ar 1024 -aa 0.05 
\end{lstlisting}
\end{minipage}
\end{minipage}
\begin{minipage}[t]{7cm}		
\begin{lstlisting}
void dielectric testmat
0
0
5 0.5 0.85 0.5 1.4 0

\end{lstlisting}
\end{minipage}

\index{ glass pane ! clear ! solid }
\index{ prism ! geometric model }
Based on model \ref{1-1-dielectric1}, including the sun.
The sun patch on the floor in above image is only found by the algorithms in this simple case of a parallelepiped volume made of dielectric.
Note the absence of a spot of direct sunlight at the ceiling which would be expected due to reflection at the lower edge of the glass block.
This is a simple example of the more general problem that the \rad numerical engine does not calculate {\em caustics}.\\
The same problems and limits apply to more complex shapes of dielectric material. Relevant applications that can {\em not} be modelled this way
include prismatic glazings or light pipes.\\
To include the redirection of the direct sun, the {\em photon map} extension of \rad is the most general approach. For a
workaround, see the specific {\tt prism} type in \ref{1-7-prism-sun}, or the more general {\tt genBSDF} pre-processor.\\
Note on the sun-patch on the floor: Illuminance levels exceed the scale used for the false-colour lines, which is set at a fixed maximum in these
comparisons. So, no false-colour lines are displayed in or around the sun patch, that is {\em not} a problem of the rendering engine.

{\bf Pro:} Fully supported with non-sun sky light and eye rays\footnote{Note: A double image of the outside tree, caused by multiple internal
reflections, is actually calculated, but isn't visible in above image due to its much lower brightness compared to the sky background.}.'

{\bf Con.:} 
The algorithms do not calculate the refraction and reflection of the direct sun in general (see \ref{1-4-dielectric2-texture} for a more
complex case). Currently, there are two workarounds available: The photon-map extension to \rad or the {\tt genBSDF} pre-processor.\\
Additionally, no scattering at front or back side and no scattering inside glass are included in the {\tt dielectric} material. Wavelength dependent refraction
is only marginally supported. so there are no chromatic effects.\\

\vfill

\newpage

\setlength{\parskip}{0.8ex}
\subsection{Solid glass window with prismatic surface, sky with sun
}

\begin{tabbing}
Objective:	\={\bf Model room with a solid block of glass with prismatic structure, sunny sky
}\\
\>\\
Scene:		\>{\bf Window is material type {\tt dielectric} combined with {\tt texfunc}
}\\
\end{tabbing}

\vspace*{-2ex}
\includegraphics[width=15cm]{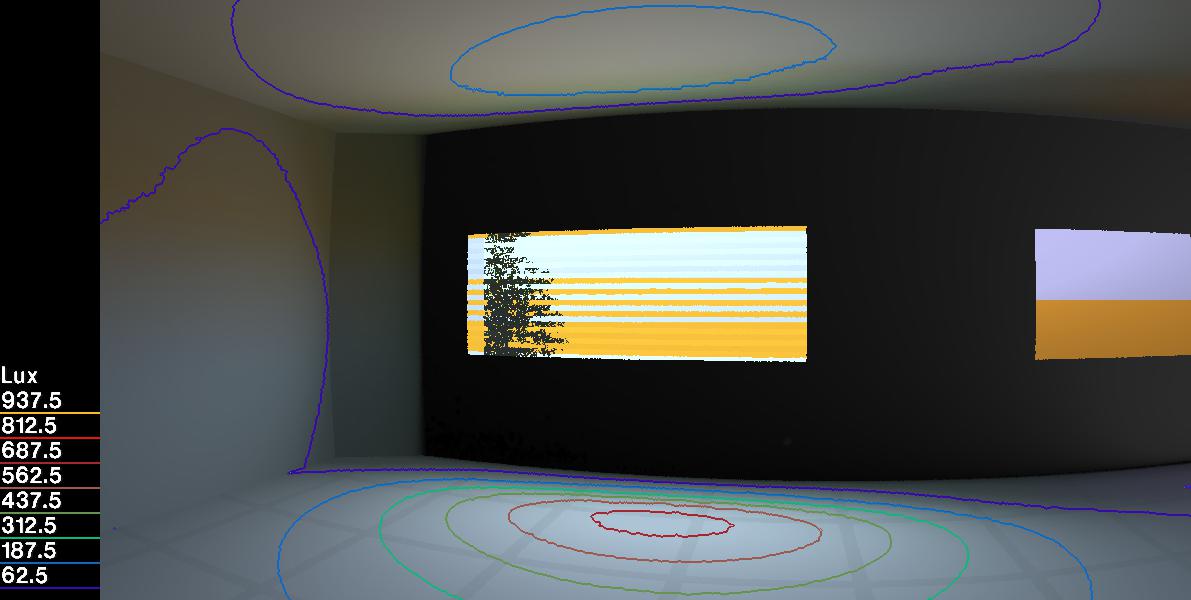}
\label{1-4-dielectric2-texture}

\lstset{ 
	backgroundcolor=\color{mygray},
	basicstyle=\ttfamily\footnotesize,
	breaklines=true,
	prebreak=\textbackslash,
	breakatwhitespace=true,
	showspaces=false,
	frame=lines,
}
\vspace*{-2ex}
\begin{minipage}[t]{8cm}		
\begin{minipage}[t]{7.4cm}		
\begin{lstlisting}
gensky 04 12 10.5CEST -a 48 -o -7.5 +i 
\end{lstlisting}
\end{minipage}\\
\begin{minipage}[t]{7.4cm}		
\begin{lstlisting}
rpict -ab 1 -ad 20000 -as 5000 -ar 1024 -aa 0.05 
\end{lstlisting}
\end{minipage}
\end{minipage}
\begin{minipage}[t]{7cm}		
\begin{lstlisting}
void texfunc tf
4 surf_Nx surf_Ny surf_Nz file1.cal
0
0
tf dielectric testmat
0
0
5 0.5 0.85 0.5 1.4 0

\end{lstlisting}
\end{minipage}

\index{ glass pane ! clear ! solid }
\index{ redirecting ! prism }
\index{ prism ! modelled by texture }
Based on model in \ref{1-2-dielectric-direct}, a perturbation of the surface normal is added. This is called a {\tt texture} in \rad.
The geometric surface itself remains flat, but when value and direction of a ray are calculated at the surface, a perturbed surface normal is
used. This perturbation is given as vector field in a user defined function file.\\
For daylighting, this could be used as an compact approximation to materials with a structured surface, as long as light-paths do not
involve multiple reflections inside the structure. Some prismatic glazings (see \ref{1-7-prism-sun}) can be modelled this way.
In this example, the size of the prisms is deliberately chosen large enough to see the structure easily.
However, this model should be applied to the direct view of the window or non-sun part of skylight only (see below). Irradiance levels
behind such a material are wrong for direct sun light, if no preprocessing step is used.

{\bf Pro:} Supported with non-sun sky light and eye rays.

{\bf Con.:} 
Direct sun light is {\em not} redirected, note missing sun spots ({\em caustics}) in image.
The algorithms do not calculate the refraction and reflection of direct light with a general {\tt dielectric} volume.
Currently, there are two workarounds: The photon-map extension to \rad or the {\tt genBSDF} pre-processor.\\
Any explicit model of the small-scale geometry of a redirecting element is prone to {\em aliasing}, caused by under-sampling the structure.
Thus, for some types of redirecting elements, the {\tt prism} model (\ref{1-7-prism-sun}) is the better alternative.\\
No scattering at front or back side and no scattering inside glass are included in the {\tt dielectric} material. Wavelength dependent refraction
is only marginally supported, so there are no chromatic effects.\\

\vfill

\newpage

\setlength{\parskip}{0.8ex}
\subsection{Micro prism window, sky with sun
}

\begin{tabbing}
Objective:	\={\bf Model room with ideal prismatic glazing, sunny sky
}\\
\>\\
Scene:		\>{\bf Window is material type {\tt prism2}
}\\
\end{tabbing}

\vspace*{-2ex}
\includegraphics[width=15cm]{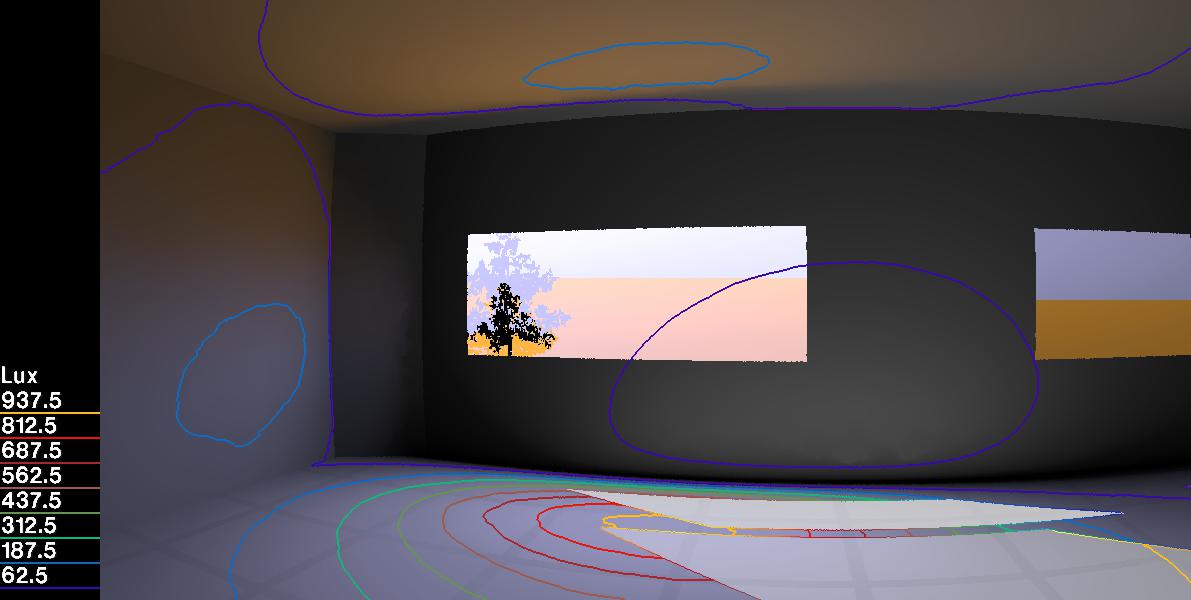}
\label{1-7-prism-sun}

\lstset{ 
	backgroundcolor=\color{mygray},
	basicstyle=\ttfamily\footnotesize,
	breaklines=true,
	prebreak=\textbackslash,
	breakatwhitespace=true,
	showspaces=false,
	frame=lines,
}
\vspace*{-2ex}
\begin{minipage}[t]{8cm}		
\begin{minipage}[t]{7.4cm}		
\begin{lstlisting}
gensky 04 12 10.5CEST -a 48 -o -7.5 +i 
\end{lstlisting}
\end{minipage}\\
\begin{minipage}[t]{7.4cm}		
\begin{lstlisting}
rpict -ab 1 -ad 20000 -ar 512 -dr 1 
\end{lstlisting}
\end{minipage}
\end{minipage}
\begin{minipage}[t]{7cm}		
\begin{lstlisting}
void prism2 testmat
9 coef1 dx1 dy1 dz1
  coef2 dx2 dy2 dz2
  prism.cal 
0
4 1.7 0.002 0.04 0.01
\end{lstlisting}
\end{minipage}

\index{ prism ! BSDF model }
Models a window pane made of prismatic light redirecting elements.
The daylighting concept itself is still a topic of recent research \cite{whang:12}.
Documentation of this add-on to \rad at Swiss EPFL, LESO group, around 1992 is comparatively sparse \cite{compagnon:93}.
See {\tt prism.cal} in the standard Radiance distribution for a starting point.\\
When using {\tt prism.cal}, supplied in the \rad package, dimensions of the prisms are given as four parameters: index of
refraction, thickness of prism triangle, height of upper side, height of lower side.\\
The {\tt prism2} material itself is not limited to prismatic glazing, it can be used to approximate any material with a maximum of two BSDF peaks.
Technically, the sun causes creation of two virtual ({\tt secondary}) light sources, whose position is calculated from the user supplied
function file.

{\bf Pro:} This material is fully supported by all light calculations in \rad: direct rays using {\tt secondary light sources} (note {\tt -dr 1} option for {\tt rpict}), ambient rays (see colour
tint at ceiling) and eye rays (double image of outside tree).
Full control over positions of maximal 2 delta peak in the BSDF, specifying directions of redirected light.
The usage of {\tt secondary sources} saves rendering time and allows a sharp redirection of direct sun light.
Calculation times are faster than the more elaborate {\tt transfunc} (see \ref{3-2-transfunc1-sun}) or {\tt BSDF} material (see \ref{3-5-bsdf1}).

{\bf Con.:} 
Model does not include scattering since this BSDF model is limited to delta peaks (two maximum). However, scattering plays an important role
in glare analysis. In fact, real materials are sincerely limited by unwanted scatter if they that are in view of occupants.
Scattering also smoothes the indoor illuminance levels. Conclusively, this material is limited in its applicability to model real, measured
materials.\\
Technically, function file for the new rays must be symmetric for incident/outgoing directions (BSDF symmetry due to reversible of optical
paths), otherwise this materials does not work correctly.

\vfill

%
%
\section{non-redirecting, forward scattering materials}
This is probably the class with the highest number of real materials. The BSDF of such materials consists of bell-shaped functions around the forward
direction.

This includes modelling of materials:
For example:
\begin{itemize}
\item	frosted glass
\item	etched or sand-blasted glass panes
\item	fabrics, shades, paper screens (e.g. Japanese shoji screens), translucent glass as a first approximation (see below)
\end{itemize}
The BSDF of some, but not all, forward scattering materials can be approximated by a Gaussian distribution around the forward direction. This makes
the \rad {\tt trans} material a natural candidate for a model.

However, it is worth remembering that less materials show a Gaussian BSDF than is commonly assumed \cite{bme:2011}.  Even if the BSDF is
fitted well by a Gaussian curve for a particular incident angle,  variations of the Gaussian parameters with incident angle are found
frequently (e.g. most fabrics used for shading) \cite{bme:2011}.
This is {\em not} included in the usual model, and these deviations between a Gaussian model and measured data lead to errors in simulation results
(see \ref{2-0-trans1} for details).

Since the BSDF is symmetric in the incident and outgoing direction, a scattering material decreases the view to the outside as well.  Even
small amounts of scattering will decrease visual perception to the outside, which is increasingly blurred. A hazy or cloudy visual appearance
is caused by an overlaid small scattering at large angles of bright light. The visual impact depend on the type of scattering (shape of the BSDF) and optional
patterns in the material (e.g. woven fabric).

\newpage

\setlength{\parskip}{0.8ex}
\subsection{Simple ideal diffuse window, sky without sun
}

\begin{tabbing}
Objective:	\={\bf Model an ideal diffuse glass in a window, non-sun sky
}\\
\>\\
Scene:		\>{\bf Window is modelled as a polygon of material type {\tt trans} with {\tt tspec=0}
}\\
\end{tabbing}

\vspace*{-2ex}
\includegraphics[width=15cm]{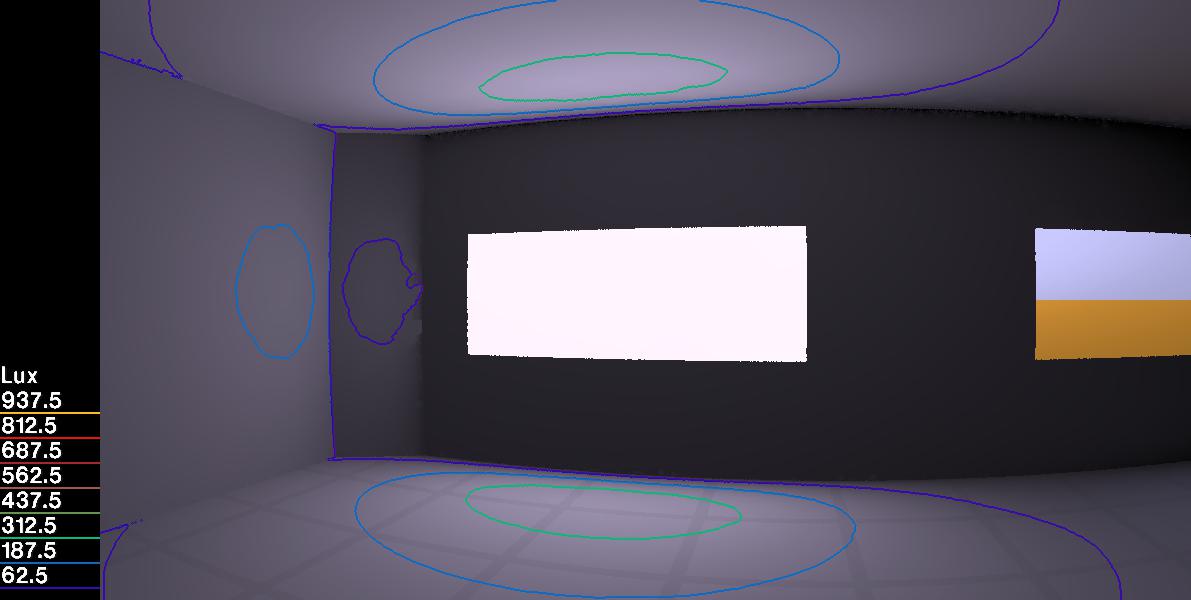}
\label{2-0-diffus1}

\lstset{ 
	backgroundcolor=\color{mygray},
	basicstyle=\ttfamily\footnotesize,
	breaklines=true,
	prebreak=\textbackslash,
	breakatwhitespace=true,
	showspaces=false,
	frame=lines,
}
\vspace*{-2ex}
\begin{minipage}[t]{8cm}		
\begin{minipage}[t]{7.4cm}		
\begin{lstlisting}
gensky 04 12 10.5CEST -a 48 -o -7.5 -i 
\end{lstlisting}
\end{minipage}\\
\begin{minipage}[t]{7.4cm}		
\begin{lstlisting}
rpict -ab 2 -ad 20000 -ar 512 
\end{lstlisting}
\end{minipage}
\end{minipage}
\begin{minipage}[t]{7cm}		
\begin{lstlisting}
void trans testmat
0
0
7  0.9 0.9 0.9  0  0.1 0.6 0
\end{lstlisting}
\end{minipage}

\index{ scattering ! Lambertian }
\index{ Lambertian scattering }
Contribution of the non-sun part of the sky at an interior point is calculated by ambient rays that hit the window where they cause a second
ambient calculation which collects light from the outside sky. 
Eye rays hitting the window are handled the same way as ambient rays: They are 100\% scattered at the window and there is zero vision through
the material (compare to \ref{2-0-trans1}).\\
The material is 10\% absorbing (first three parameter) and from the 90\% rest, 60\% are transmitted (6th parameter) (54\% transmitted total) in this example. 
There is no bluish/brownish tint on the inside surfaces, since the colour of the sky and ground get completely mixed at the diffuse window.
The indoor illuminance is symmetric to the centre of the window for the same reason.\\
Note that two ambient bounces ({\tt -ab 2}) are needed to get any light into the room: The second bounce collects light at the outside of
the window from the sky model.

{\bf Pro:} This material is fully supported by all light calculations in \rad, which in this case means that the non-sun part of the sky (a more or less diffuse sky) is correctly taken into
account.\\
The {\tt trans} material has only a few parameters, which are documented in \cite{radiance-mat}.
Setting the {\tt tspec} parameter to zero makes the material ideally diffusing ({\em Lambertian scatterer}, constant BSDF).

{\bf Con.:} 
A translucent material that scatters in an ideal diffuse way is extremely rare, even in laboratories, and certainly in practice. All
existing window materials show some forward scattering.\\
As said, this material needs one ambient bounce more than the forward scattering materials in \ref{2-0-trans1}.

\vfill

\newpage

\setlength{\parskip}{0.8ex}
\subsection{Translucent window, sky without sun
}

\begin{tabbing}
Objective:	\={\bf Model a sandblasted glass in a window, non-sun sky
}\\
\>\\
Scene:		\>{\bf Window is modelled as a polygon of material type {\tt trans}
}\\
\end{tabbing}

\vspace*{-2ex}
\includegraphics[width=15cm]{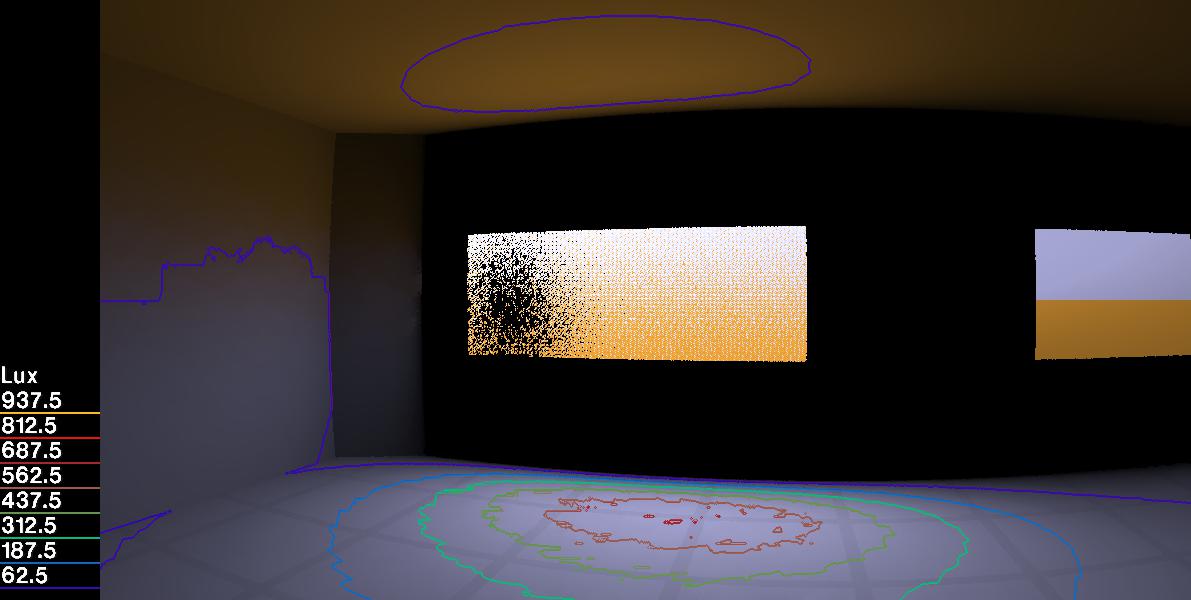}
\label{2-0-trans1}

\lstset{ 
	backgroundcolor=\color{mygray},
	basicstyle=\ttfamily\footnotesize,
	breaklines=true,
	prebreak=\textbackslash,
	breakatwhitespace=true,
	showspaces=false,
	frame=lines,
}
\vspace*{-2ex}
\begin{minipage}[t]{8cm}		
\begin{minipage}[t]{7.4cm}		
\begin{lstlisting}
gensky 04 12 10.5CEST -a 48 -o -7.5 -i 
\end{lstlisting}
\end{minipage}\\
\begin{minipage}[t]{7.4cm}		
\begin{lstlisting}
rpict -ab 1 -ad 20000 -ar 512 
\end{lstlisting}
\end{minipage}
\end{minipage}
\begin{minipage}[t]{7cm}		
\begin{lstlisting}
void trans testmat
0
0
7  0.9 0.9 0.9  0  0.1 0.6 1
\end{lstlisting}
\end{minipage}

\index{ scattering ! forward }
\index{ glass pane ! sandblasted }
\index{ trans }
Contribution of the non-sun part of the sky at an interior point is calculated by ambient rays that are transmitted and scattered at the window.
The amount of scattering depends on the roughness parameter. 
\footnote{This scattering of ambient rays are handled internally by an implicit inversion of the Gaussian BSDF model.}\\
Eye rays hitting the window are handled the same way as ambient rays: They are scattered at the window and thereby provide a blurred vision
through the material.\\
Overall hemispherical transmission is the same as in \ref{2-0-diffus1}. Note the shifted maximum compared to \ref{2-0-diffus1} due to the
asymmetric distribution of a morning sky and the forward scattering of the window.\\
The same features and limits apply to the asymmetric version, the \rad model {\tt trans2}.\index{trans2}

{\bf Pro:} 
This material is fully supported by all light calculations in \rad, which in this case means that the non-sun part of the sky is correctly taken into account.
Note that the similarly named model {\tt transfunc}, is numerically completely different (see \ref{3-1-transfunc1}).\\
The model has been applied to some materials in daylighting, e.g. Aerogels \cite{apian-ti7:94}.
A low number of parameters for {\tt trans} material makes it easy to fit them to measured data, if they are reasonably well approximated by
the Gaussian BSDF model.

{\bf Con.:} 
The shape of the {\tt trans} BSDF is fixed to a Gaussian curve: $f(x)= b e^{-a x^2}$. The user parameter control only the width and height of the
Gaussian distribution and are constant over incident angles.\\
Measurements show that many materials scatter differently \cite{bme:2011}, which introduces errors in simulation results.
Angular selective materials (including most sun-shade fabrics) are not well matched by this model, since their transmission parameters depends on the
incident direction. Other materials, notably metals, are not well matched by the Gaussian BSDF model in the first place \cite{bme:2011}.

\vfill

\newpage

\setlength{\parskip}{0.8ex}
\subsection{Translucent window: combination of transparent and diffusing glass, sky without sun
}

\begin{tabbing}
Objective:	\={\bf Model an partially edged glass in a window, non-sun sky
}\\
\>\\
Scene:		\>{\bf Window is modelled as a polygon of combined material with type {\tt mixfunc}
}\\
\end{tabbing}

\vspace*{-2ex}
\includegraphics[width=15cm]{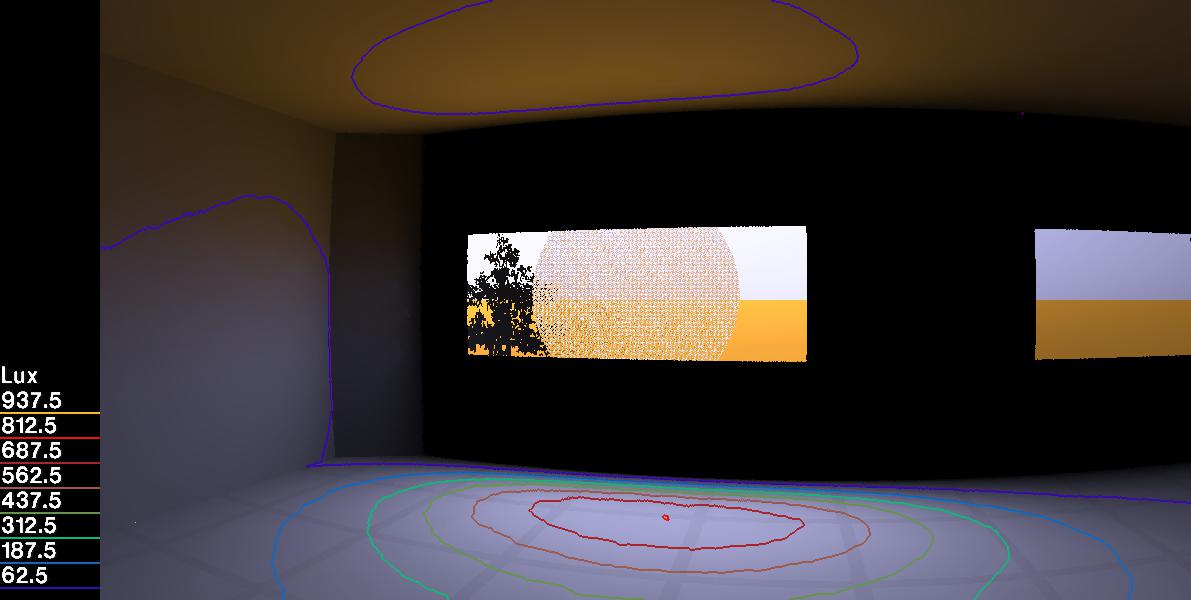}
\label{2-1-mixfunc1}

\lstset{ 
	backgroundcolor=\color{mygray},
	basicstyle=\ttfamily\footnotesize,
	breaklines=true,
	prebreak=\textbackslash,
	breakatwhitespace=true,
	showspaces=false,
	frame=lines,
}
\vspace*{-2ex}
\begin{minipage}[t]{8cm}		
\begin{minipage}[t]{7.4cm}		
\begin{lstlisting}
gensky 04 12 10.5CEST -a 48 -o -7.5 -i 
\end{lstlisting}
\end{minipage}\\
\begin{minipage}[t]{7.4cm}		
\begin{lstlisting}
rpict -ab 1 -ad 20000 -ar 512 
\end{lstlisting}
\end{minipage}
\end{minipage}
\begin{minipage}[t]{7cm}		
\begin{lstlisting}
void trans mat1
0
0
7  0.9 0.9 0.9  0  0.3 0.6 1
void glass mat2
0
0
3 0.96 0.96 0.96
void mixfunc testmat
4 mat2 mat1 val file1.cal
0
0
\end{lstlisting}
\end{minipage}

This combines the material {\tt glass} (\ref{0-2-glass1}) and {\tt trans} (\ref{2-0-trans1}) into a new material.
The mixture of the two is controlled by a function file, allowing fine-tuning: In this example the mixture depends
on the position on the window, in the same way that the glass absorption was modified in \ref{0-2-glass2}.
More complex functions are easily built.\\
Note the increase in noise at the iso contour lines, which is caused by the variation of transmittance across the window surface. A
higher number of ambient rays ({\tt ad} parameter) would be needed to get the same low noise as in \ref{2-0-diffus1}. This is a more general
problem of numerically integrating a non-constant function: The more the function fluctuates over the integration domain, the denser the
sample points have to be.

{\bf Pro:} 
Flexible method to create a new material from two base materials. Support for ambient, direct and eye rays depend on support of each base material.

{\bf Con.:} 
Due to internal control flow, \rad (up to and including the current version 4.1) does not offer the incident direction 
in the function file that controls {\tt mixfunc}. Therefor, angular selective and scattering shadings (e.g. fabric sunshades) can not be modelled this way.

\vfill

%
%
\section{redirecting, scattering materials}

The material class with seems the most relevant:
Daylighting requires redirection of light to get more even indoor illumination levels across the room, and it requires some scattering to get a
homogeneous indoor illumination without caustics. These are much the same guidelines used for reflector design in lamps.
A well-fitting description of real materials in this class needs the most general model.

This includes:
\begin{itemize}
\item	multi-walled, extruded transparent plastic sheets (clear plastics like PMMA or PC)
\item	BSDF representation of geometrically complex models of light redirecting louvres, using curved reflectors and/or scattering materials
	(as used by the \rad preprocessor {\tt genBSDF})
\item	aluminium reflector sheets with regular, structured surface (e.g. saw-tooth profile) used in light-shelfs
\item	other redirecting elements, e.g. laser-cut-panels, crazes, holographic/mircostructured elements
\end{itemize}
During the schematic design phase, the simpler models of redirecting-non-scattering materials (\ref{1-7-prism-sun}) could be sufficient.
In later stages during planning, material models are based on measured BSDF data. They should model scattering if materials show scattering,
to avoid serious glare problems, for example. That requires a model of this class.

Since the BSDF is symmetric in the incident and outgoing direction, a typical scattering and redirecting material may decrease the view to the
outside considerably, details depend on the type of scattering (shape of the BSDF). In general, the visual quality of the window itself should be
represented in the model, to estimate the view through the material and any possible discomfort caused by glare.

\newpage

\setlength{\parskip}{0.8ex}
\subsection{Redirecting+scattering window, legacy material, sky without sun
}

\begin{tabbing}
Objective:	\={\bf Model arbitrary window material, non-sun sky
}\\
\>\\
Scene:		\>{\bf Window is modelled as a polygon of material type {\tt transfunc}
}\\
\end{tabbing}

\vspace*{-2ex}
\includegraphics[width=15cm]{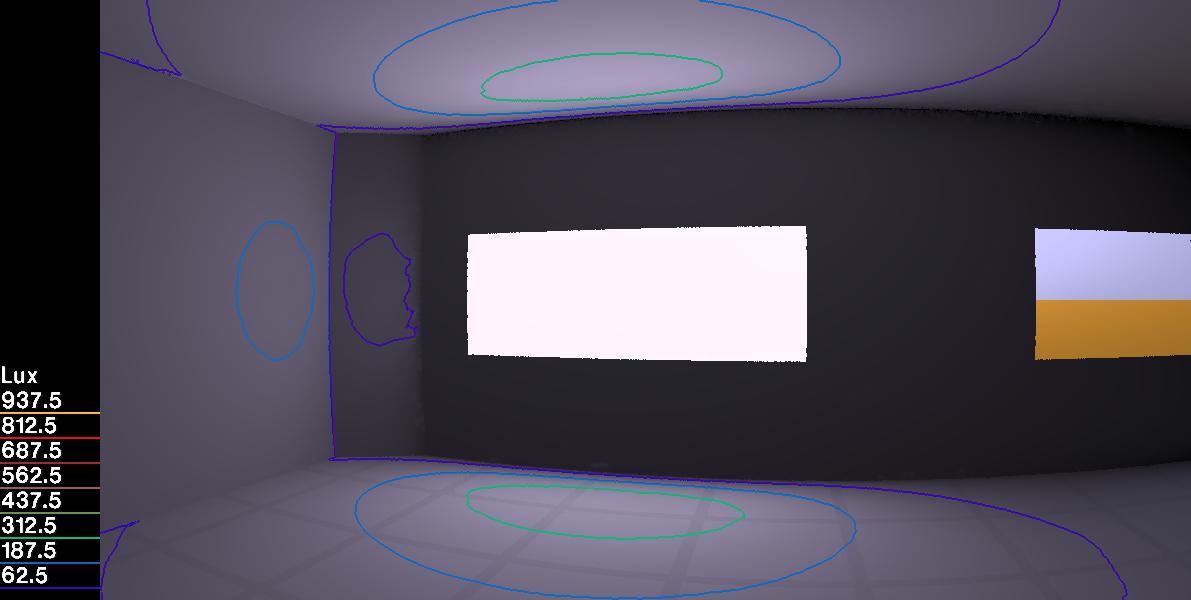}
\label{3-1-transfunc1}

\lstset{ 
	backgroundcolor=\color{mygray},
	basicstyle=\ttfamily\footnotesize,
	breaklines=true,
	prebreak=\textbackslash,
	breakatwhitespace=true,
	showspaces=false,
	frame=lines,
}
\vspace*{-2ex}
\begin{minipage}[t]{8cm}		
\begin{minipage}[t]{7.4cm}		
\begin{lstlisting}
gensky 04 12 10.5CEST -a 48 -o -7.5 -i 
\end{lstlisting}
\end{minipage}\\
\begin{minipage}[t]{7.4cm}		
\begin{lstlisting}
rpict -ab 2 -ad 20000 -ar 512 
\end{lstlisting}
\end{minipage}
\end{minipage}
\begin{minipage}[t]{7cm}		
\begin{lstlisting}
void transfunc testmat
6 sval file1.cal -rx 90 -ry 0
0
6   0.9 0.9 0.9   0 0.6 1
\end{lstlisting}
\end{minipage}

\index{ redirecting ! general }
The {\tt transfunc} model offers BSDF specification by a user-supplied function, making it one of the most general, adaptable and therefore
important material gorup\footnote{{\tt transdata}, {\tt brightfunc}, {\tt brightdata}, {\tt BRDFdata} have identical
algorithms, {\tt BRTDfunc} offers additional peaks} in \rad.
It is also probably the least commonly understood material of the program:\\
Using of {\tt transfunc} with a sun-less sky is problematic: ambient and eye rays are {\em not} transmitted or scattered at the
window. Numerically these rays are handled differently from the closely named model-based {\tt trans} (\ref{2-0-trans1}), a fact often
overlooked.
For a sun-less sky, the supplied function is not evaluated, but approximated by a BSDF constant over outgoing and incoming directions.
So the inside contribution of the window is ideal diffuse, irrespective of the supplied BSDF function
\footnote{Note that using the \rad preprocessor {\tt mkillum} does {\em not} solve this problem, since eye rays used by {\tt rtrace}, which
is called by {\tt mkillum}, show the same deficit.}. To sample the outside non-sun-sky an additional ambient calculation ({\tt ab}) is needed,
increasing calculation times.\\
The supplied function $f(\vec{x})$ is expected to be normalised: $\int^{hemisphere} f(\vec{x})\, \cos\theta_{o}\, d\vec{x} = 1$.
For a BSDF with integrals of hemispherical-hemispherical transmission and reflection $\tau_{hh}, \rho_{hh}$, the first three parameters
({\tt RGB}) should then be set to $\tau_{hh}+\rho_{hh}$ and the 5th parameter ({\tt trans}) to $\tau_{hh}/(\tau_{hh}+\rho_{hh})$.

{\bf Pro:} An arbitrary BSDF can be specified by the user using a functional language (see example in \ref{3-2-transfunc1-sun}).

{\bf Con.:} 
Approximates BSDF crudely as constant with a sun-less sky (see \ref{3-5-bsdf1} for correct result).  A {\tt transfunc} model with user supplied BSDF
function is therefore unsuitable for light simulation of indoor light levels with a sky distribution where the majority of incoming light is
not direct sun light.\\
Setting the correct parameters of {\tt transfunc}, needed to at least approximate the BSDF for a
sun-less sky, requires extra calculation of integrated values $\tau_{hh},\rho_{hh}$ from the BSDF function by the user.\\
See \ref{2-0-diffus1} for a truly diffuse material, \ref{3-2-transfunc1-sun} for a sunny sky and \ref{3-5-bsdf1} for further, newer use of function files.
\index{ mkillum } \index{transdata} \index{brightfunc} \index{brightdata} \index{BRDFdata}

\vfill

\newpage

\setlength{\parskip}{0.8ex}
\subsection{Redirecting+scattering window, legacy material, sky with sun 
}

\begin{tabbing}
Objective:	\={\bf Model arbitrary material in a window, sunny sky
}\\
\>\\
Scene:		\>{\bf Window is modelled as a polygon of material type {\tt transfunc}
}\\
\end{tabbing}

\vspace*{-2ex}
\includegraphics[width=15cm]{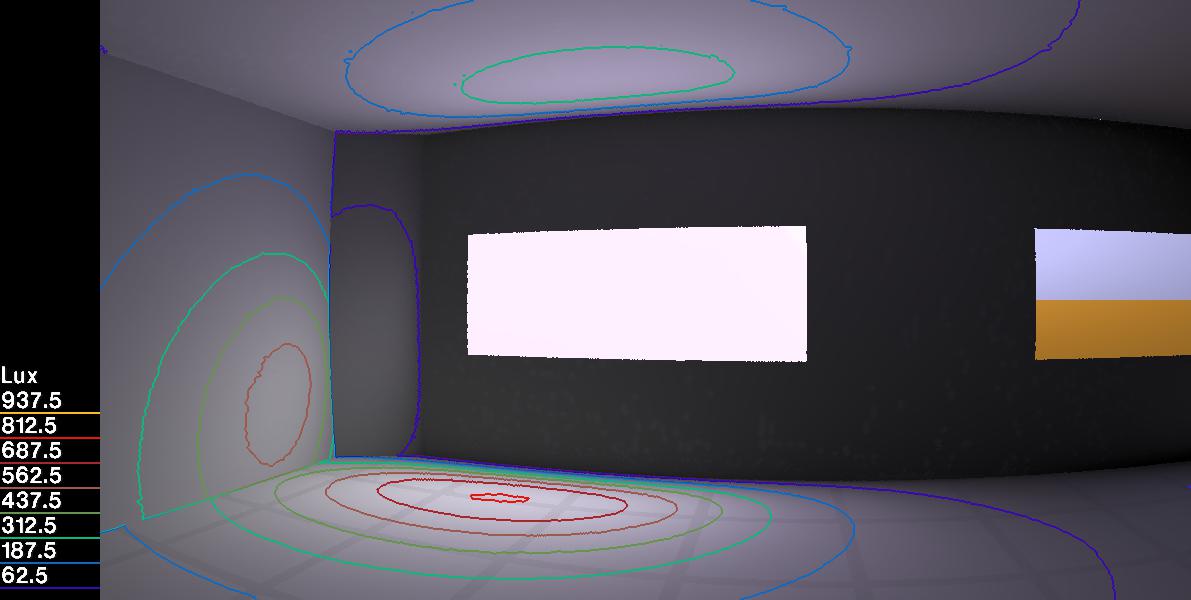}
\label{3-2-transfunc1-sun}

\lstset{ 
	backgroundcolor=\color{mygray},
	basicstyle=\ttfamily\footnotesize,
	breaklines=true,
	prebreak=\textbackslash,
	breakatwhitespace=true,
	showspaces=false,
	frame=lines,
}
\vspace*{-2ex}
\begin{minipage}[t]{8cm}		
\begin{minipage}[t]{7.4cm}		
\begin{lstlisting}
gensky 04 12 10.5CEST -a 48 -o -7.5 +i 
\end{lstlisting}
\end{minipage}\\
\begin{minipage}[t]{7.4cm}		
\begin{lstlisting}
rpict -ab 2 -ad 20000 -as 5000 -ar 1024 -aa 0.05 
\end{lstlisting}
\end{minipage}
\end{minipage}
\begin{minipage}[t]{7cm}		
\begin{lstlisting}
void transfunc testmat
6 sval file1.cal -rx 90 -ry 0
0
6   0.9 0.9 0.9   0 0.6 1
\end{lstlisting}
\end{minipage}

Same material as in \ref{3-1-transfunc1}.
Contribution of any part of the sky at an interior point is calculated by ambient rays, which trigger a direct calculation at the
window, which then uses the user supplied BSDF for the direct sun. Eye rays hitting the window are handled the same way.
The function for both cases:
\begin{lstlisting}
SQ(x)=x*x;
our_bsdf(x,y,z,Dx,Dy,Dz)= exp(-SQ( 2*Acos( +x*Dx - y*Dy - z*Dz))) / (0.3*3.1415);
sval(x,y,z,o)= (1/arg(1)) * (1/arg(5)) * our_bsdf(x,y,z,-Dx,-Dy,-Dz);
\end{lstlisting}
This BSDF scatters side-wards, leading to higher illuminance levels left of the window.
\footnote{The redirection is cause by the plus sign at the {\tt x*Dx} term. Note the rotation of the coordinates in the fourth parameter to
{\tt transfunc}:  x-axis is along the window, y-axis is vertically up and z-axis points outside.}
Compared to \ref{3-1-transfunc1}, indoor light levels are caused by redirection of direct sun light.
\footnote{
Note scaling of function {\tt our\_bsdf} by the first and fifth parameter: {\tt transfunc} expects a normalised function
(\ref{3-1-transfunc1}).}
See also \ref{3-6-bsdf1-sun} for {\tt BSDF} material.

{\bf Pro:} Specifically with this \rad material, the incident direction is available in the function file. It is a powerful method when based
on measured BSDF data, since fitting parameters of a function to measured BSDF results in a compact, general BSDF model. \cite{apian:95}.\\
Use {\tt BRTDfunc} for additional peaks in forward and mirror direction (although no arbitrary peak direction can be given in the
functionfile then).

{\bf Con.:} 
The user defined BSDF is used for direct sources (direct sun light) only. This can lead to significant errors for indoor illumination
calculations, depending on the sky conditions (see \ref{3-1-transfunc1}).\\
Note that calculation of non-scattering redirection ({\tt prism} described in \ref{1-7-prism-sun}) is at least an order of magnitude faster.
The other materials in this group ({\tt transdata}, {\tt brightfunc}, {\tt brightdata}, {\tt BRTDfunc}) have identical algorithms and
therefore identical limits.
\index{ fabrics } \index{ transdata } \index{ brightfunc }

\vfill

\newpage

\setlength{\parskip}{0.8ex}
\subsection{Redirecting+scattering window, new material, sky without sun
}

\begin{tabbing}
Objective:	\={\bf Model arbitrary window material, non-sun sky
}\\
\>\\
Scene:		\>{\bf Window is modelled as a polygon of material type {\tt BSDF}
}\\
\end{tabbing}

\vspace*{-2ex}
\includegraphics[width=15cm]{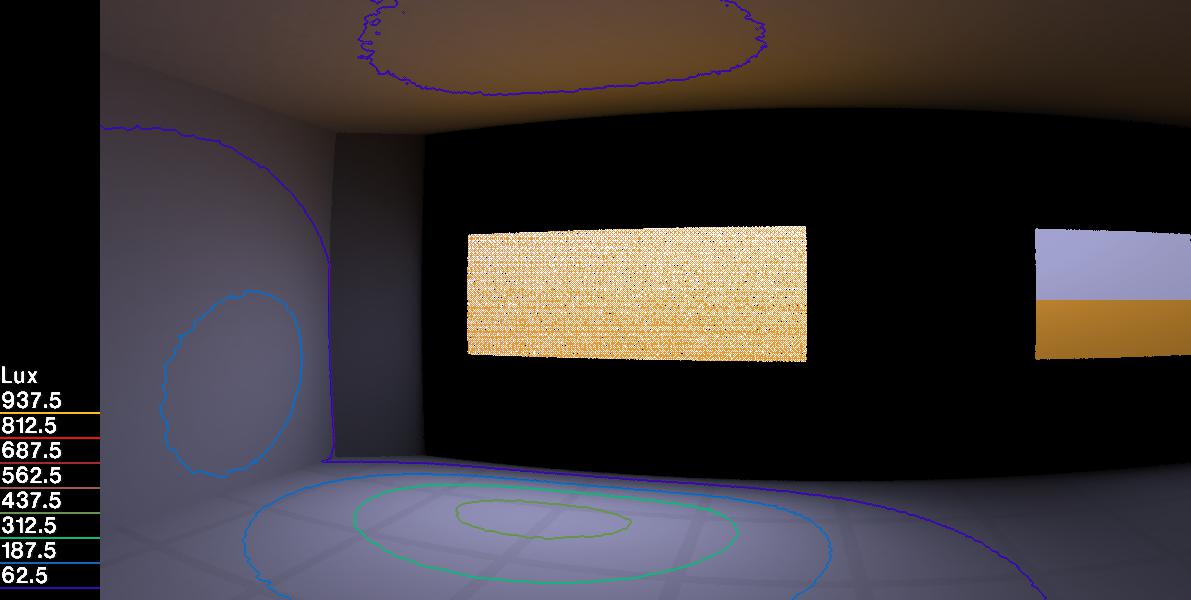}
\label{3-5-bsdf1}

\lstset{ 
	backgroundcolor=\color{mygray},
	basicstyle=\ttfamily\footnotesize,
	breaklines=true,
	prebreak=\textbackslash,
	breakatwhitespace=true,
	showspaces=false,
	frame=lines,
}
\vspace*{-2ex}
\begin{minipage}[t]{8cm}		
\begin{minipage}[t]{7.4cm}		
\begin{lstlisting}
gensky 04 12 10.5CEST -a 48 -o -7.5 -i 
\end{lstlisting}
\end{minipage}\\
\begin{minipage}[t]{7.4cm}		
\begin{lstlisting}
rpict -ab 1 -ad 20000 -as 5000 -ar 1024 -aa 0.05 
\end{lstlisting}
\end{minipage}
\end{minipage}
\begin{minipage}[t]{7cm}		
\begin{lstlisting}
void BSDF testmat
8 0 file1.xml 0 1 0 . -rx 90 
0
0
\end{lstlisting}
\end{minipage}

The material type {\tt BSDF} uses an auxiliary file with discrete BSDF data values, in XML (extensible marker language) format.
In this particular example, the file is generated by the new \rad {\tt bsdf2ttree} converter, added in 2013. It reads a function file
with a BSDF model and generates the XML file (see footnote in \ref{3-6-bsdf1-sun} for the command).\\
In this example, the BSDF function is the same as in \ref{3-1-transfunc1}.
Note the shifted, and now correctly calculated, illuminance distribution on the floor and lower illuminance levels at the ceiling,  compared to \ref{3-1-transfunc1}.
The colour tints are also correctly calculated with the colours of the outside sources.\\
Generally, usage of a pre-canned BSDF file in XML notation requires less setup by the user, since more auxiliary data can be included within
the XML structure.  Using XML, the specifications of its format are well defined and allow a broader application outside \rad .
These XML files can be generated by any external program, based on an abstract, stand-alone, well-defined XML syntax definition
\footnote{\url{See, for example, http://en.wikipedia.org/wiki/Xml}}. Therefor, this material offers a generic way of getting BSDF data into \rad.

{\bf Pro:} This material is fully supported by all light calculations in \rad.
Therefore this model calculates correctly, especially compared to the older {\tt transfunc} model (\ref{3-1-transfunc1}).
Rendering times are substantially less then in \ref{3-1-transfunc1}, since it uses one ambient calculation less.

{\bf Con.:} 
Currently the XML format does not include definitions of delta-peaks in the BSDF, so no sharp shadows and see-through. Secondly, it lists
the BSDF as discrete values, potentially limiting angular resolution of narrow scattering, see also \ref{3-6-bsdf1-sun}.\\
Even with one ambient calculation less, the number of ambient rays (see \ref{ambientrays}) has to be higher than for simpler examples
\ref{0-0-void1} to get the same level of stochastic noise.

\vfill

\newpage

\setlength{\parskip}{0.8ex}
\subsection{Redirecting+scattering window, new material, sky with sun 
}

\begin{tabbing}
Objective:	\={\bf Model arbitrary window material, sunny sky
}\\
\>\\
Scene:		\>{\bf Window is modelled as a polygon of material type {\tt BSDF}
}\\
\end{tabbing}

\vspace*{-2ex}
\includegraphics[width=15cm]{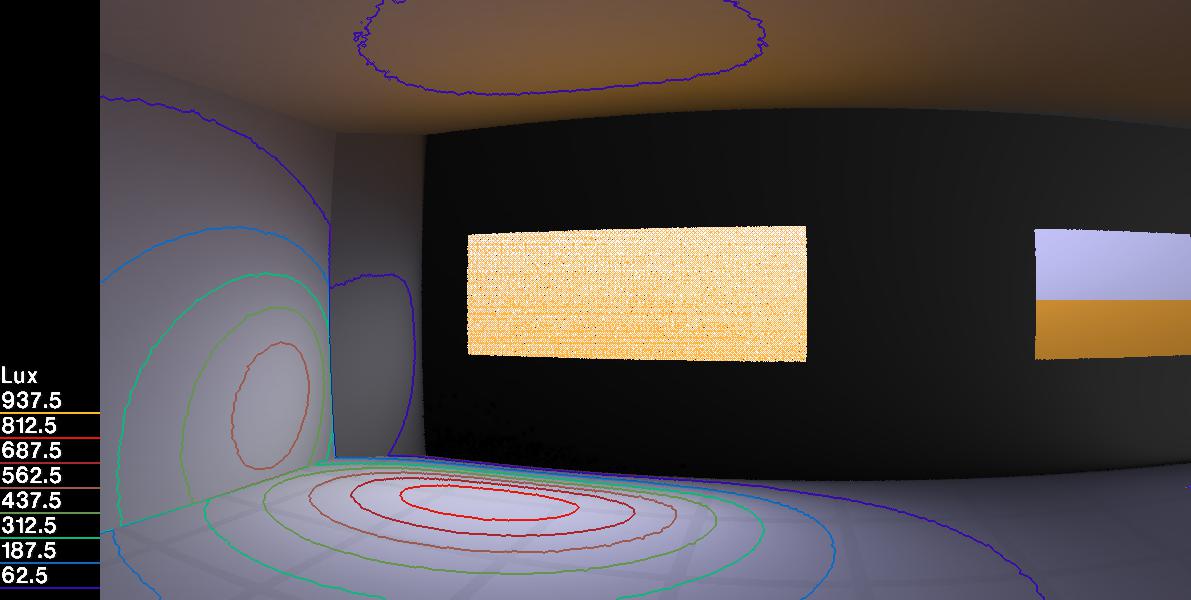}
\label{3-6-bsdf1-sun}

\lstset{ 
	backgroundcolor=\color{mygray},
	basicstyle=\ttfamily\footnotesize,
	breaklines=true,
	prebreak=\textbackslash,
	breakatwhitespace=true,
	showspaces=false,
	frame=lines,
}
\vspace*{-2ex}
\begin{minipage}[t]{8cm}		
\begin{minipage}[t]{7.4cm}		
\begin{lstlisting}
gensky 04 12 10.5CEST -a 48 -o -7.5 +i 
\end{lstlisting}
\end{minipage}\\
\begin{minipage}[t]{7.4cm}		
\begin{lstlisting}
rpict -ab 1 -ad 20000 -as 5000 -ar 1024 -aa 0.05 
\end{lstlisting}
\end{minipage}
\end{minipage}
\begin{minipage}[t]{7cm}		
\begin{lstlisting}
void BSDF testmat
8 0 file1.xml 0 1 0 . -rx 90 
0
0
\end{lstlisting}
\end{minipage}

\index{ BSDF ! material }
See \ref{3-5-bsdf1} for an introduction to the {\tt BSDF} model.
\footnote{command line: bsdf2ttree +backward +forward -t4 -g 6 -f rayinit.cal -f file1.cal our\_bsdf $>$ file1.xml .
To use the same coordinates as in \ref{3-2-transfunc1-sun}, note the inversion of the D-vector in the definition of {\tt sval}
in\ref{3-2-transfunc1-sun} and the {\em up} direction along a (0,1,0) vector. Note, while the simple function file has a size of 134 bytes,
the XML data needs 100MB in this case. On the other hand, the XML file size is mostly uncorrelated to the complexity of the generating function.} .
The main benefit of discrete data is the possibility of an ad-hoc inversion of a user supplied BSDF during calculation, which is the technical
reason for the added functionality of the BSDF material. Effectively, this allows a user defined BSDF function to be used for all sky
conditions, a substantial improvement over previous models (\ref{3-1-transfunc1}).
\\
In this case, the BSDF function is the same as in \ref{3-1-transfunc1}. Note the difference in illumination distribution on the floor,
compared to \ref{3-2-transfunc1-sun}.

{\bf Pro:} This material is fully supported by all light calculations in \rad. See the example \ref{3-5-bsdf1} for advantages.
This new material offers more precise calculations at less numerical overhead, and without the need to extra calculate integrals 
$\tau_{hh}, \rho_{hh}$ in order to set parameters, as previously needed \ref{3-1-transfunc1}.

{\bf Con.:} 
The current XML format does not offer a way to describe peaks in the discrete BSDF-XML data.
The angular resolution, intrinsic to sampling a BSDF function, limits the following: Visual see-through quality, sharp shadows and sharp glare
patterns when viewed directly.\\
It does offer a workaround for the problem of see-through quality, by automatically switching to a second model
for the visual appearance of the window itself. This is used by the supplementary program {\tt genBSDF}, which generates BSDF-XML file from
a geometrical model of a shading device.
An equivalent tool to handle measured BSDF data, including interpolation and identifications/extraction of delta-peaks components, is not yet available.\\
The material could also be combined with older models, such as {\tt prism} (\ref{1-7-prism-sun}) or {\tt transfunc} (\ref{3-1-transfunc1})
to enhance representation of the view-through quality.

\vfill

\section{Conclusions}



Numerical models for light redirecting materials have been a feature of \rad since the start.
Recent enhancements to its material models (e.g. {\tt BSDF}) expand its simulation
capabilities of special window materials. 
This text provides a summarising snapshot of the current capabilities, both the older ones (sections \ref{0-2-glass2} to \ref{3-2-transfunc1-sun})
and newest one (sections \ref{3-5-bsdf1} to \ref{3-6-bsdf1-sun}) and may help to bridge the gap between the capabilities of the numerical
engine and application of models for complex fenestration.

The suggested four categories of window materials show the correlation between typical applications of daylighting fenestration, numerical
procedures in \rad and its applicable material models.


Simulating daylighting with \rad consists of mainly four parts: geometry, window material, sky model and simulation parameters. The last
three are interrelated. User interfaces on top of the \rad core may shield details from the user by auto-selecting
standard combinations known to work. However, more detailed understanding is required for detailed analysis and modelling of new materials.

From the perspective of material modelling, any simulation program with less choice of material models than \rad is unlikely worth
considering when dealing with complex fenestration, since these materials exists, are applicable in practice and need simulation for their
well guided use.

\section{Outlook}

Three useful future developments may be concluded:
Firstly, adding a notation for BSDF delta peaks (already partly implemented, see \ref{1-7-prism-sun}) to function files.
Numerically, they have to be treated separately, in post-processing measurements and for visual quality of a window material and shadows
by direct sun-light. Coding in the function file would keep all characteristics in one place. 
Function files could be embedded in the XML framework, providing a homogeneous, powerful and centralised BSDF description.

Secondly, the long-favoured model-less-interpolation-algorithm (MOLIA) will get measured BSDF data more easily into the XML BSDF schema, without
a manually devised functional BSDF model. A key element in this algorithm will be an ''intelligent'' interpolation between incident angles,
which preserves the topological characteristics of arbitrary BSDF shapes.

Thirdly, it may be useful to homogenise and extended the core raytracing engine: Incorporating the photon-map or Metropolis algorithms
would add to this tool. It would benefit material modelling and overall-use, algorithms should ''number crunch'' different daylighting
scenarios with less parameters. The current selection of ''correct'' material models and simulation parameters for a specific material
could be made easier for a user by enhancing the underlying simulation engine.

\section{Acknowledgements}
Many thanks to Greg Ward for his friendly and very professional support over the years, and to Eleanor Lee of LBNL for 
coordinating the \rad development over the last few years.

\pagebreak[6]

\section{Disclaimer, referencing this paper, review options }

The contents has been carefully cross-checked and is believed to be accurate material. No warranty what so ever is implied. Use any information
at your own risk.

This work is freely available at no charge for download at \url{http://www.pab.eu/docs/}.
It has been filed at \url{http://arxiv.org/}, so it is best to use its DOI number {\em to-be-filled-in} if you want to reference this text.\\
For options of an anonymous peer-review see \url{http://code.google.com/p/gpeerreview/} and
\url{http://en.wikipedia.org/wiki/Open_peer_review}. \\
For the general problem with some established journals and their policies, see
\url{http://www.guardian.co.uk/science/peer-review-scientific-publishing} and other sides.\\
Feedback by email to the \href{http://www.radiance-online.org/}{\rad mailing list} or to the \href{mailto:papers@pab.eu}{author} is also appreciated.


%
%
\bibliographystyle{alpha}

\section{Bibliography}
\bibliography{iselib,lbl,misc,optics,arch,delaunay,rendering,sig1} 

\end{document}